\pdfoutput=1

\documentclass[acmsmall]{acmart}

\renewcommand\footnotetextcopyrightpermission[1]{} 
\AtBeginDocument{%
  }
    
\setcopyright{acmlicensed}
\copyrightyear{2024}
\acmYear{2024}
\acmDOI{XXXXXXX.XXXXXXX}

\acmJournal{PACMPL}
\acmVolume{8}
\acmNumber{OOPSLA2}
\acmArticle{1}

\usepackage{booktabs}   
\usepackage{subcaption} 

\usepackage{amsmath}    
\usepackage{amsthm}
\usepackage{textcomp} 
\usepackage{caption}
\usepackage{listings}
\usepackage{xcolor}
\usepackage[xcolor]{changebar}
\usepackage{graphicx}
\usepackage{xspace}
\usepackage{algorithm}
\usepackage[noend]{algpseudocode}
\usepackage{multirow}
\usepackage[export]{adjustbox}
\usepackage{enumitem}
\usepackage{float}
\usepackage{tikz}
\usepackage{makecell}
\usepackage{siunitx}
\usepackage{pifont}
\usepackage{hyphenat}
\usepackage{wrapfig}
\usepackage{tikz}

\usetikzlibrary{calc,positioning}

\makeatletter
\newcommand{\myfsize}{\f@size pt}
\makeatother

\newfloat{algorithm}{t}{lop}

\algnewcommand{\IIf}[1]{\State\algorithmicif\ #1\ \algorithmicthen}
\algnewcommand{\EndIIf}{\unskip\ \algorithmicend\ \algorithmicif}

\newcommand{\metacomment}[3]%
{{\color{#1}#2:\@#3}}

\newcommand{\heading}[1]{\vspace{4pt}\textbf{#1}\enspace}
\newcommand{\papertitle}

\newcommand{\sparse}[1]{{\textbf{\textit{#1}}}}

\newcommand{\brackets}[1]{{\langle{#1}\rangle}}
\DeclareMathOperator{\Indices}{Indices}
\DeclareMathOperator{\Schedules}{sch}
\DeclareMathOperator{\nnz}{nnz}
\DeclareMathOperator{\loopfuse}{\textbf{loopfuse}}

\DeclareMathOperator{\reorder}{\textbf{reorder}}

\DeclareMathOperator{\sparsity}{sparsity}
\DeclareMathOperator{\loopC}{loop}
\newcommand{\idx}{\ensuremath{\text{\textit{idx}}}}

\newcommand{\etal}{\textit{et al.}~}
\newcommand{\eg}{\textit{e.g.,}~}
\newcommand{\ie}{\textit{i.e.,}~}

\algnewcommand\algorithmicforeach{\textbf{for each}}
\algdef{S}[FOR]{ForEach}[1]{\algorithmicforeach\ #1\ \algorithmicdo}

\definecolor{lstgreen}{rgb}{0,0.5,0}
\newcommand\encircle[1]{%
  \tikz[baseline=(X.base)] 
    \node (X) [draw, shape=circle, inner sep=0] {\strut #1};}

\newcommand{\veryshortarrow}[1][4pt]{\hspace{-1.5pt}\mathrel{%
  \hbox{\rule[\dimexpr\fontdimen22\textfont2-.2pt\relax]{#1}{0.6pt}}%
    \mkern-4mu\hbox{\usefont{U}{lasy}{m}{n}\symbol{41}}}\hspace{-1.5pt}}

\graphicspath{ {./images/} }

\newcommand{\system}{SparseAuto\xspace}
\newcommand{\code}[1]{\texttt{#1}} 

\begin{document}

\title[SparseAuto]{SparseAuto: An Auto-Scheduler for Sparse Tensor Computations Using Recursive Loop Nest Restructuring}

\author{Adhitha Dias}
\orcid{0000-0003-3500-7547}
\affiliation{
  \department{Electrical and Computer Engineering}
  \institution{Purdue University}
  \streetaddress{610 Purdue Mall}
  \city{West Lafayette}
  \state{IN}
  \country{USA}
  \postcode{47906}
}
\email{kadhitha@purdue.edu}

\author{Logan Anderson}
\affiliation{
  \department{Electrical and Computer Engineering}
  \institution{Purdue University}
  \streetaddress{610 Purdue Mall}
  \city{West Lafayette}
  \state{IN}
  \country{USA}
  \postcode{47906}
}
\email{anderslt@purdue.edu}

\author{Kirshanthan Sundararajah}
\orcid{0000-0001-6384-062X}
\affiliation{
  \department{Computer Science}
  \institution{Virginia Tech}
  \city{Blacksburg}
  \state{VA}
  \country{USA}
  \postcode{24061}
}
\email{kirshanthans@vt.edu}

\author{Artem Pelenitsyn}
\orcid{0000-0001-8334-8106}
\affiliation{
  \department{Electrical and Computer Engineering}
  \institution{Purdue University}
  \streetaddress{610 Purdue Mall}
  \city{West Lafayette}
  \state{IN}
  \country{USA}
  \postcode{47906}
}
\email{apelenit@purdue.edu}

\author{Milind Kulkarni}
\orcid{0000-0001-6827-345X}
\affiliation{
  \department{Electrical and Computer Engineering}
  \institution{Purdue University}
  \streetaddress{610 Purdue Mall}
  \city{West Lafayette}
  \state{IN}
  \country{USA}
  \postcode{47906}
}
\email{milind@purdue.edu}

\begin{abstract}
  Automated code generation and performance enhancements for sparse tensor algebra have become essential in many real-world applications, such as quantum computing, physical simulations, computational chemistry, and machine learning. 
General sparse tensor algebra compilers are not always versatile enough to generate {\em asymptotically optimal code} for sparse tensor contractions.  
This paper shows how to generate asymptotically better schedules for complex sparse tensor expressions using kernel fission and fusion.
We present generalized loop restructuring transformations to reduce asymptotic time complexity and memory footprint.

Furthermore, we present an auto-scheduler that uses a partially ordered set (poset)-based cost model that uses both time and auxiliary memory complexities to prune the search space of schedules. 
In addition, we highlight the use of Satisfiability Module Theory (SMT) solvers in sparse auto-schedulers to approximate the Pareto frontier of better schedules to the smallest number of possible schedules, with user-defined constraints available at compile-time. 
Finally, we show that our auto-scheduler can select better-performing schedules and generate code for them. 
Our results show that the auto-scheduler provided schedules achieve orders-of-magnitude speedup compared to the code generated by the Tensor Algebra Compiler (TACO) for several computations on different real-world tensors.

\end{abstract}



\maketitle

\section{Introduction}\label{introduction}
 
Tensor contractions are used in many real-world applications such as physical simulations, machine learning, computational chemistry, and quantum computing~\cite{tce2003,simquant,tncbook,Kossaifi_2017_CVPR_Workshops,ran2017review}.
When most of the values in these tensors become zero, it is advantageous to use compact data formats to store only the non-zero values, and such tensors are known as {\em sparse tensors}.
We can exploit the zeros in sparse tensors by skipping the computations over them (\ie \code{x + 0 = x} and \code{x * 0 = 0}).
Many important applications such as Graph Neural Networks (GNN), Physical Simulation, and Quantum Chemistry~\cite{hamilton2017inductive,featgraph,fusedmm} make use of sparse tensor contractions.

Due to the {\em compressed data formats} that store the sparse tensors, their contractions are realized as {\em non-affine} loop nests, where bounds depend on the input, and accesses are indirect.
The non-affine loop nests of sparse tensor contractions prevent us from directly applying classical affine loop transformation frameworks to reduce load imbalances and bad locality for performance enhancement.
This challenge has given rise to specialized compilers for sparse tensor computations~\cite{taco,aartbik:93,compiler_in_mlir,michelle2015,pingali97,senanayake:2020:scheduling,aart22,kjolstad:2018:workspaces} and various abstractions for the schedule --- realization of computation (\eg loop structure, parallelization etc.) --- to separate it from the computation.
Schedule abstractions (\S~\ref{scheduling_primitives}) make it convenient to realize a plethora of ways to materialize a computation using transformations such as loop reordering, loop fusion/fission, loop tiling, loop parallelization, etc.

Choosing a better-performing schedule for a sparse tensor contraction is not straightforward. Therefore, it is more challenging than finding a schedule for its dense counterpart, which is realized as {\em affine} loop nests (\ie There exist well-studied analytical cost models of schedules and machines for dense tensor computations).
The schedule selection heavily depends on sparse tensor inputs (number of non-zero values and sparsity structure), making it difficult to pick a performant one for sparse tensor computations.
Hence, the simplest method to evaluate the cost of a schedule is to execute it on a given machine using the provided sparse tensor inputs to measure the time it takes to finish the execution.
The sheer number of schedules makes it an arduous time-consuming process; therefore, it is not practical to execute all schedules to find the best one.
Also, it is important to note that there may not be a single best schedule for all sparse tensor inputs and machines, and some schedules may be {\em asymptotically} better than others.

The challenge in finding a performant schedule for sparse tensor contractions arises due to two main factors: {\em vast space of schedules} and {\em heavy dependency on sparse tensor inputs}.
We provide a systematic way to {\em completely} explore the vast space of schedule at compile-time rooted in transformations (\ie loop reordering and loop/kernel fusion/fission), which makes it convenient to realize the schedule.
The exploration of the schedule space is augmented with {\em machine-independent} pruning strategies and {\em symbolic sparse tensor input attributes} at compile-time to filter most of the schedules and keep a handful of schedules to be evaluated at run-time with {\em machine-dependent} parameters and {\em concrete sparse tensor input attributes} to select a performant schedule.
As there are asymptotically superior schedules, the pruning strategy encompasses comparing schedules for {\em both} time and auxiliary memory complexity, which depends on the attributes of sparse tensor inputs.
To the best of our knowledge, prior work does not optimize for both time and auxiliary memory complexity~\cite{willowahrens,solomonik}.

Consider this example of sparse tensor times matrix contraction: $A_{lmn} = \sum_{ijk}\>{\sparse B_{ijk}}\>C_{il}\>D_{jm}\>E_{kn}$~\footnote{Bold face letters denote sparse tensors.}. 
This computation can be expressed using a simple linear loop nest with a time complexity of $O(\nnz({\sparse B_{ijk}})LMN)$. 
Alternatively, the contraction can be expressed as $T_{ijn} = \sum_{k}\>B_{ijk}\>E_{kn}$ and $A_{lmn} = \sum_{ij}\>T_{ijn}\>C_{il}\>D_{jm}$ -- two separate computations with a total time complexity of $O(\nnz({\sparse B_{ijk}})N + IJNLM)$ and a dense temporary $T$. 
Another schedule can be obtained from the observation that the outer loops of the first computation (producer) can be fused with the second computation (consumer) (\ie loop fusion). 
This schedule reduces the overall time complexity to $O(N(\nnz({\sparse B_{ijk}}) + \nnz({\sparse B_{ij}})LM))$~\footnote{$\nnz({\sparse B_{ij}})$ refers to iterating only the first two levels in the indexing arrays of ${\sparse B_{ij}}$ without visiting the third dimension $k$.} and a scalar temporary, which is asymptotically superior to both of the previous schedules in time and memory complexity. 
The last schedule has a branching loop structure (\ie {\em imperfectly nested} loop nest) that is different from the other two schedules, which have simple loop structures (\ie {\em perfectly nested} loop nests).
However, the last schedule dominates the other two schedules in terms of both time and memory complexity (\S~\ref{overview}).
Therefore, to explore schedules with multi-level branching loop structures, which are of asymptotically superior time and auxiliary memory complexity, we introduce the {\em extended representation of branched iteration graphs}~\cite{sparselnr} and a new {\em scheduling directive} to realize such schedules (\S~\ref{detailed_design}).
Furthermore, we explore the schedule space of a given sparse tensor computation and present strategies based on {\em partially ordered sets} (posets) that can be combined with user-defined constraints at compile-time to prune the schedule space (\S~\ref{autoscheduler}). 
Contributions of this paper are as follows:

\begin{description}
\item[Recursive extension of branched iteration graph] We generalize the branched iteration graph (BIR) representation of SparseLNR~\cite{sparselnr} to support schedules with multiple levels of imperfectly nested loops and new {\em scheduling primitives} to realize the schedules by recursively applying loop/kernel fusion/fission with loop reorder.

\item[Complete schedule space exploration] We provide a strategy to explore the schedules of a given sparse tensor contraction {\em guaranteed} to cover the {\em complete} space of schedules with loop structures, including multi-level branching (\ie multiple levels of imperfectly nested loops), attainable using loop/kernel fusion/fission.

\item[Novel auto-scheduler] We introduce a novel {\em poset-based} auto-scheduler to prune the space of schedules to create a Pareto frontier wrt. {\em both} time and auxiliary memory complexity. We use a Satisfiability Modulo Theory (SMT) solver to compare the symbolic time and memory complexity with {\em user-defined} constraints.
\end{description}

The rest of the paper is organized as follows. 
We provide the necessary background in Section~\ref{background} and in Section~\ref{overview}, we motivate the problem.
The multi-level branched iteration graph and the scheduling primitives are introduced in Section~\ref{detailed_design}.
We discuss schedule exploration and selection in Section~\ref{autoscheduler}.
Evaluation of our auto-scheduler is presented in Section~\ref{evaluation}.
We conclude the paper in Section~\ref{conclusion} with a discussion.

\section{Background}\label{background}

This section discusses the necessary background on sparse tensor access constraints, tensor index notation, iteration graph representation, and scheduling primitives to understand the challenge in auto-scheduling for sparse tensor contraction.

\subsection{Sparse Tensor Access Constraints}\label{sparse_tensors}

There are several compressed data formats used to store sparse tensors: Compressed Sparse Row (CSR), Sorted Coordinate (Sorted COO), Compressed Sparse Fiber (CSF), etc., to name a few.
These formats are abstracted by {\em level format}~\cite{chou}, a tree structure that shows the order in which index arrays must be traversed to retrieve an element.
The {\em sparse tensor access constraints} are imposed by the order of access of the index arrays in compressed data formats.
For example, if ${\sparse A_{ij}}$ is in CSR format, the row index should be traversed to get to the column index, which results in a dependency between $i$ and $j$, the indices traverse rows and columns of ${\sparse A}$, respectively.
Therefore, the loops $i$ and $j$ belong to cannot be freely reordered.
TACO Format Abstraction~\cite{chou} describes the level formats in detail.

\subsection{Tensor Index Notation for Tensor Contractions}\label{tensor_contraction}

The notation that describes tensor contraction operations is based on the Einstein Summation (Einsum) convention. 
This notational convention implies summation over a set of repeated indices. 
For example, the expression $X(i,k) = A(i,j)\cdot B(j,k)$ implies summation over the repeated index $j$ and equivalent to the standard mathematical notation $A_{ik} = \sum\nolimits_{j} B_{ij} C_{jk}$\footnote{This is the matrix-matrix multiply operation.}. 
We use both these notations interchangeably in the text. 
Since this computation can be performed using a simple linear triply nested loop, its iteration time complexity is $O(IJK)$, where $I$, $J$, and $K$ are the loop bounds. 
If ${\sparse B}$ is sparse, then the iteration time complexity is $O(\nnz({\sparse B_{ij}})K)$, where $\nnz$ is the number of non-zero elements.

\subsection{Iteration Graph}\label{iteration_graph}

\begin{figure*}[t]
    \centering
    \begin{subfigure}{0.3\linewidth}
        \centering
        \includegraphics[width=0.4\linewidth]{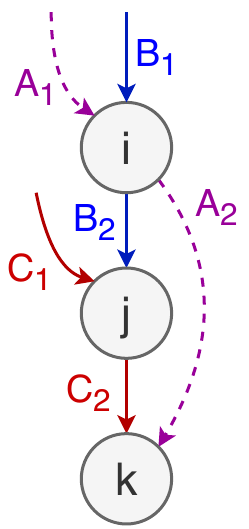}
        \caption{}
        \label{fig:spmm_iteration_graph}
    \end{subfigure}
    \begin{subfigure}{0.58\linewidth}
        \begin{lstlisting}[language=C++,tabsize=2]
for (int32_t i = 0; i < B1_dimension; i++) {
    for (int32_t jB = B2_pos[i]; jB < B2_pos[(i + 1)]; jB++) {
        int32_t j = B2_crd[jB];
        for (int32_t k = 0; k < C2_dimension; k++) {
            int32_t kA = i * A2_dimension + k;
            int32_t kC = j * C2_dimension + k;
            A_vals[kA] = A_vals[kA] + B_vals[jB] * C_vals[kC];
        }
    }
}
        \end{lstlisting}
        \vspace{-1.5em}
        \caption{}
        \label{fig:spmm_code}
    \end{subfigure}
    \vspace{-0.5em}
    \caption{An example of an iteration graph for sparse matrix-matrix multiplication and corresponding code.}
    \label{fig:spmm-graph-and-code}
\end{figure*}

Consider the example sparse matrix-matrix multiplication (SpMM), $A_{ik} = \sum_{k}\>{\sparse B_{ij}}\>C_{jk}$. 
An iterator that iterates through all of $i$, $j$, and $k$ can read each value of ${\sparse B_{ij}}$, $C_{jk}$, multiply each value sharing the same $j$, and store the result in $A_{ik}$. 
An example iteration graph is shown in Figure~\ref{fig:spmm_iteration_graph}, and this internal representation (IR) is used to generate code in Figure~\ref{fig:spmm_code}. 
The nodes in the iteration graph represent indices in the Einsum notation. 
This is an acyclic graph where the edges represent the dimensions of tensors and how they map to indices. 
Since ${\sparse B}$ is sparse, $B_{1}$ and $B_{2}$, incidents on indices $i$ and $j$ must not change the order, and other indices can appear in any order as they traverse dense tensors (\eg $C_{1}$ and $C_{2}$). 
No other sparse tensor access constraints (\S~\ref{sparse_tensors}) are imposed. 
TACO~\cite{taco} describes the concept of iteration graphs in detail.

\subsection{Scheduling Primitives}\label{scheduling_primitives}

A schedule describes one way of realizing a computation, and multiple schedules can realize the same computation. 
For example, we can change the loop order in Figure~\ref{fig:spmm-graph-and-code} to get the order $i,k,j$ instead of $i,j,k$. 
TACO~\cite{taco} and Sparse Iteration Space Framework~\cite{senanayake:2020:scheduling} describe the importance of abstractions to separate the computation from the schedule. 
The algorithmic and scheduling languages describe the computation and schedule, respectively, and scheduling primitives form the scheduling language. 
Some of the scheduling primitives are as follows: {\em reorder} to reorder the loops; {\em split} to split a loop for tiling; {\em collapse} to collapse one loop onto another; {\em parallelize} and {\em vectorize} for parallel execution. 
TACO-Workspaces~\cite{kjolstad:2018:workspaces} introduces {\em precompute} to add dense intermediaries to schedules.
In Section~\ref{detailed_design}, we introduce a new scheduling primitive called $\loopfuse$, which can produce loop nests with branched loops (\ie imperfectly nested loops) combined with the $\reorder$ directive.  

\section{Overview}\label{overview}

It may not be straightforward to decide whether to apply transformations across multiple kernels. 
The decision depends both on the iteration complexity of the final loop nests and the working set sizes.\footnote{Iteration complexity refers to the number of total iterations in a loop nest required to complete the computation. For example, iteration complexity of the kernel in Figure~\ref{fig:a-tensor-contraction-kernel} is $LMN\cdot(\nnz(B))$}
If the working set sizes are small and fit into the cache then, it is better to use the version with lower iteration complexity. 
Otherwise, it is better to use the schedule with lower auxiliary memory. 
Hence, an auto-scheduler that only looks at the iteration complexity or only the auxiliary memory complexity may choose the wrong schedule as the final output or prune a good schedule from the search space in the process.

  \begin{figure*}[t!]
    \begin{adjustbox}{minipage=\linewidth,scale=0.9}
    \centering
      \begin{subfigure}[t]{0.47\textwidth}
        \begin{lstlisting}[basicstyle=\small, gobble=8, tabsize=2, showtabs=false, showstringspaces=false]
          for perm(l,m,n,i,j_pos,k_pos):
            A(l,m,n)+=B(i,j,k)*C(i,l)*D(j,m)*E(k,n)
        \end{lstlisting}
        \caption[For the list of figures]{$ {A_{lmn}} = \sum\nolimits_{ijk} {\sparse B_{ijk}}\cdot C_{il}\cdot D_{jm}\cdot E_{kn} $\\Time: $LMN\cdot \nnz(B)$,\quad Memory: $0$\\Loop Depth: $6$,\quad Memory Depth: $0$}
        \label{fig:a-tensor-contraction-kernel}
      \end{subfigure}%
      \vskip\baselineskip
      \begin{subfigure}[t]{0.47\textwidth}
        \begin{lstlisting}[basicstyle=\small, gobble=8, tabsize=2, showstringspaces=false]
          for perm(m, l):
            T<k> = 0
            for i, j_pos, k_pos:
              T(k) += B(i,j,k)*C(i,l)*D(j,m)
            for perm(n, k):
              A(l,m,n) += T(k)*E(k,n)
        \end{lstlisting}
        \caption[For the list of figures]{$ {A_{lmn}} = \sum\nolimits_{k} (\sum\nolimits_{ij} {\sparse B_{ijk}}\cdot C_{il}\cdot D_{jm})\cdot E_{kn} $\\Time: $LM\cdot(\nnz(B)+NK)$,\quad Memory: $K$\\Loop Depth: $5$,\quad Memory Depth: $1$}
        \label{fig:b-tensor-contraction-kernel}
      \end{subfigure}%
      \hspace{10pt}
      \begin{subfigure}[t]{0.47\textwidth}
        \begin{lstlisting}[basicstyle=\small, gobble=8, tabsize=2, showstringspaces=false]
          for l:
            T<k,j> = 0
            for i, j_pos, k_pos:
              T(k,j) += B(i,j,k)*C(i,l)
            for perm(j, m, k, n):
              A(l,m,n) += T(k,j)*D(j,m)*E(k,n)
        \end{lstlisting}
        \caption[For the list of figures]{$ {A_{lmn}} = \sum\nolimits_{jk} (\sum\nolimits_{i}{{\sparse B_{ijk}}\cdot C_{il}})\cdot D_{jm}\cdot E_{kn} $\\Time: $L\cdot(\nnz(B)+JMKN)$,\quad Memory: $KJ$\\Loop Depth: $5$,\quad Memory Depth: $2$}
        \label{fig:e-tensor-contraction-kernel}
      \end{subfigure}
      \vskip\baselineskip
      \begin{subfigure}[t]{0.47\textwidth}
        \vspace{-10pt}
        \begin{lstlisting}[basicstyle=\small, gobble=8, tabsize=2, showstringspaces=false]
          for l:
            T<k,j> = 0
            for i, j_pos, k_pos:
              T(k,j) += B(i,j,k)*C(i,l)
            for perm(m, k):
              t = 0
              for j:
                t += T(k,j)*D(j,m)
              for n:
                A(l,m,n) += t*E(k,n)
        \end{lstlisting}
        \caption[For the list of figures]{$ {A_{lmn}} = \sum\nolimits_{k} (\sum\nolimits_{j} (\sum\nolimits_{i} {\sparse B_{ijk}}\cdot C_{il})\cdot D_{jm})\cdot E_{kn} $\\Time: $L\cdot(\nnz(B)+MK(J+N))$,\quad Memory: $KJ$\\Loop Depth: $4$,\quad Memory Depth: $2$}
        \label{fig:c-tensor-contraction-kernel}
      \end{subfigure}%
      \hspace{10pt}%
      \begin{subfigure}[t]{0.47\textwidth}
        \begin{lstlisting}[basicstyle=\small, gobble=8, tabsize=2, showstringspaces=false]
          for l:
            T<j,k> = 0
            for i, j_pos, k_pos:
              T(k,j) += B(i,j,k)*C(i,l)
            T<m,k> = 0
            for perm(j, m, k):
              T(m,k) += T(j,k)*D(j,m)
            for perm(m, k, n):
              A(l,m,n) += T(m,k)*E(k,n)
        \end{lstlisting}
        \caption[For the list of figures]{$ {A_{lmn}} = \sum\nolimits_{k} (\sum\nolimits_{j} (\sum\nolimits_{i} {\sparse B_{ijk}}\cdot C_{il})\cdot D_{jm})\cdot E_{kn} $\\Time: $L\cdot(\nnz(B)+MK(J+N))$,\quad Memory: $KJ+MK$\\Loop Depth: $4$,\quad Memory Depth: $2$}
        \label{fig:d-tensor-contraction-kernel}
        \Description[C++ code of the kernel fused sampled dense-dense matrix multiplication kernel and sparse matrix multiplication kernel]{<long description>}
      \end{subfigure}
    \end{adjustbox}
    \caption{Different schedules of executing $ {A_{lmn}} = \sum\nolimits_{ijk} {\sparse B_{ijk}}\cdot C_{il}\cdot D_{jm}\cdot E_{kn} $. Here, the code snippet~\ref{fig:a-tensor-contraction-kernel} has a perfectly nested loop structure while all the other code snippets has a nested loop structure. Here, $j\_pos$ refers to the non-affine loop associated with the index $j$. The loop $j\_pos$ is non-affine because ${\sparse B_{ij}}$ is sparse. The code snippets~\ref{fig:b-tensor-contraction-kernel} and~\ref{fig:e-tensor-contraction-kernel} has one level of branching whereas the code snippets~\ref{fig:c-tensor-contraction-kernel} and~\ref{fig:d-tensor-contraction-kernel} has a branch nesting depth of two.}
    \label{fig:tensor-contraction-kernel}
  \end{figure*}

\subsection{Motivating Example}\label{motivating_example}

There may be many schedules to perform a tensor contraction, and which one to
choose depends on your viewpoint.
Consider the following example involving a sparse tensor $\sparse B$: 
\[A(l,m,n) = {\sparse B(i,j,k)} * C(i,l) * D(j,m) * E(k,n)\]

Figure~\ref{fig:a-tensor-contraction-kernel} refers to performing the computation using a simple loop nest of depth 6.
The same computation can be written as in figures~\ref{fig:b-tensor-contraction-kernel},~\ref{fig:e-tensor-contraction-kernel},~\ref{fig:c-tensor-contraction-kernel}, and ~\ref{fig:d-tensor-contraction-kernel} with branching loop nests of depth 4 or 5. 
In this section, we will discuss the performance of these different schedules. 
We will evaluate all the schedules with the same loop structure, but with different index ordering and report the best one. 
For the loop structure in Figure~\ref{fig:c-tensor-contraction-kernel}, we will evaluate both the inner loop order of $m, k$ and $k, m$. 
Similarly, for Figure~\ref{fig:b-tensor-contraction-kernel}, we will evaluate four different loop orders, two of them by interchanging the inner loops $n, k$ and two of them by interchanging the outer loops $l, m$.

From the asymptotic time complexity viewpoint, an auto\hyp{}scheduler might lean towards pruning Schedule~\ref{fig:a-tensor-contraction-kernel}.
This is due to its loop nesting depth of 6 and time complexity of $O(\nnz(B_{IJK})LMN)$, in contrast to the schedule in Figure~\ref{fig:b-tensor-contraction-kernel} with a loop nesting depth of 5 and asymptotic time complexity of $O(\nnz(B_{IJK})LM + LMNK)$ or the schedule in Figure~\ref{fig:c-tensor-contraction-kernel} with a loop nesting depth of 4 and asymptotic time complexity of $O(\nnz(B_{IJK})L + LMK(J+N))$.
Notably, the schedule in Figure~\ref{fig:d-tensor-contraction-kernel} has the same asymptotic time complexity as the schedule in Figure~\ref{fig:c-tensor-contraction-kernel}, while the asymptotic time complexity of the schedule in Figure~\ref{fig:e-tensor-contraction-kernel} is $O(\nnz(B_{IJK})L + LJMKN)$.

\begin{wrapfigure}{r}{0.4\textwidth}
    \centering
    \includegraphics[width=0.35\textwidth, trim=20pt 580pt 350pt 20pt, clip]{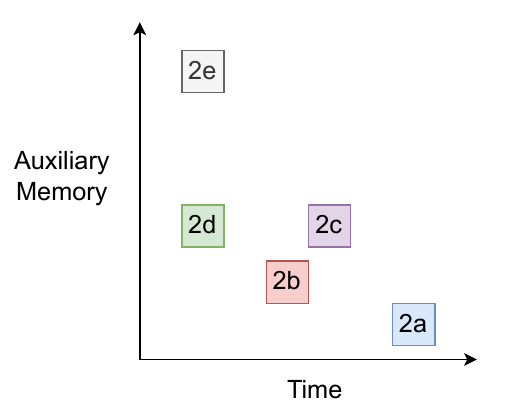}
    \caption{Placement of schedules based on asymptotic time vs. auxiliary memory complexities.}
    \label{placement.pdf}
    \vspace{-1.0em}
  \end{wrapfigure}

These schedules can be placed on a asymptotic time complexity vs. auxiliary memory complexity space plot as shown in Figure~\ref{placement.pdf}, relative to each other.

From the perspective of asymptotic time complexity, an auto-scheduler might favor either the schedule in Figure~\ref{fig:c-tensor-contraction-kernel} or Figure~\ref{fig:d-tensor-contraction-kernel}, both having a loop depth of 4, the lowest among the five schedules in Figure~\ref{fig:tensor-contraction-kernel}.
Comparing these two schedules, Figure~\ref{fig:c-tensor-contraction-kernel} uses one 2D auxiliary memory for storing intermediate results between branched loop nests, while Figure~\ref{fig:d-tensor-contraction-kernel} uses a 2D and a 1D auxiliary memory.
Consequently, the former has lower memory complexity than the latter. In summary, the schedule in Figure~\ref{fig:c-tensor-contraction-kernel} dominates Figure~\ref{fig:d-tensor-contraction-kernel}, as both schedules share the same asymptotic time complexity, but the former is better in terms of auxiliary memory complexity.

Comparing the schedules in Figures~\ref{fig:c-tensor-contraction-kernel} and~\ref{fig:e-tensor-contraction-kernel} from the asymptotic memory complexity perspective, both exhibit an auxiliary memory complexity of $O(KJ)$. The time complexity of the former, Figure~\ref{fig:c-tensor-contraction-kernel}, is superior with $O(J+N)$ being better than $O(JN)$ for larger values of $J$ and $N$.
Consequently, Figure~\ref{fig:c-tensor-contraction-kernel} dominates Figure~\ref{fig:e-tensor-contraction-kernel}. For the sake of brevity, comparisons involving Figures~\ref{fig:d-tensor-contraction-kernel} and~\ref{fig:e-tensor-contraction-kernel} with other schedules are omitted in the following paragraphs.

Consider the comparison of the schedules in Figures~\ref{fig:a-tensor-contraction-kernel}, \ref{fig:b-tensor-contraction-kernel}, and~\ref{fig:c-tensor-contraction-kernel} when the bounds change in the range as follows; $1 \leq I \leq 1800$, $1 \leq J \leq 1600$, $400 \leq K \leq 4000$, $8 \leq L \leq 256$, $8 \leq M \leq 256$, $8 \leq N \leq 256$, and $0.001 \leq \sparsity(B) \leq 0.01$.
Note that the schedule in Figure~\ref{fig:a-tensor-contraction-kernel} dominates both the schedules in Figures~\ref{fig:b-tensor-contraction-kernel} and~\ref{fig:c-tensor-contraction-kernel} in terms of the auxiliary memory usage because no auxiliary memory is used in the Schedule~\ref{fig:a-tensor-contraction-kernel}.
Although the loop depth is four for the schedule in Figure~\ref{fig:c-tensor-contraction-kernel}, within the given ranges of bounds and the sparsity of $B$, we cannot claim that it is the best in all cases.
Let us look at some cases by changing the loop bounds and sparsity for tensor $B$.
The evaluation configuration is explained in Section~\ref{evaluation}.

\heading{Case \textcircled{1}: Lowest loop depth schedule (Figure~\ref{fig:c-tensor-contraction-kernel}) is the best}
\label{experiment_1} 
In this case, we set the loop bounds for the schedules in Figure~\ref{fig:tensor-contraction-kernel} to specific values: $I = 1800$, $J = 800$, $K = 1000$, $L = 64$, $M = 16$, $N = 32$\footnote{$I, J, K, L, M,$ and  $N$ are the loop bounds of loops with indices $i, j, k, l, m$, and $n$, respectively.}, and $\sparsity(B) = 0.08$.
Under these conditions, the iteration time complexities follow the inequality
$\Phi(d)~\footnote{$\Phi(x)$ refers to the iteration time complexity of the schedule in Figure2$x$ for concrete bounds in the given Case.} < 7.5*\Phi(d)  \thickapprox \Phi(b) < 237.5 * \Phi(d) \thickapprox \Phi(a)$.
The corresponding execution times follow the inequality
$\Psi(d)~\footnote{$\Psi(x)$ refers to the execution time of the schedule in Figure 2$x$.} = 2.48s < \Psi(b) = 6.26s < \Psi(a) = 32.40s$.
The schedule in Figure~\ref{fig:c-tensor-contraction-kernel} exhibits the lowest loop depth and iteration time complexity.
An auto-scheduler that factors in loop depth could choose the best schedule in this case. 

\heading{Case \textcircled{2}: Effect of the size of auxiliary memory}
\label{experiment_2} 
Adjusting the loop bounds to $J = 1600$, $K = 2000$ and $\sparsity(B) = 0.02$ while maintaining other loop bounds as in the previous example, the schedule in Figure~\ref{fig:c-tensor-contraction-kernel} now incurs an auxiliary temporary memory requirement of $12.21MB$ (compared to $3.05MB$ in Case \textcircled{1}).
This consumes more than $50\%$ of the last-level cache (LLC). 
The iteration time complexities follow the inequality: $\Phi(d) < 3*\Phi(d) \thickapprox \Phi(b) < 92.5 * \Phi(d) \thickapprox \Phi(a)$.
Execution times for the schedules follow the inequality, $\Psi(b) = 8.60s < \Psi(d) = 10.20s < \Psi(a) = 33.51s$.
The schedule in Figure~\ref{fig:c-tensor-contraction-kernel}, with the minimum loop depth, exhibits the lowest iteration time complexity as in the previous example, but the schedule in Figure~\ref{fig:b-tensor-contraction-kernel}, with a loop depth of 5, performs better.  
It is evident from this case that a good auto-scheduler must consider the sizes of the auxiliary memory arrays used in the computation. 
Consequently, an auto-scheduler solely reliant on loop depth would fail in this scenario. 

\heading{Case \textcircled{3}: Highest loop \& lowest memory depth schedule (Figure~\ref{fig:a-tensor-contraction-kernel}) is the best}
\label{experiment_3}
Setting loop bounds and sparsity as $I = 1$, $J = 200$, $K = 4000$, $L = 256$, $M = 200$, $N = 196$, and $\sparsity(B) = 0.002$, the execution times of the schedules in Figure~\ref{fig:a-tensor-contraction-kernel}, Figure~\ref{fig:b-tensor-contraction-kernel}, and Figure~\ref{fig:c-tensor-contraction-kernel} follows the inequality: $\Psi(a) = 4.9 ms < \Psi(b) = 9.1 ms < \Psi(d) = 282.5 ms$. This scenario is an example where the schedule with the highest loop depth (Schedule $a$) executes the fastest. An auto-scheduler that factors in loop depth would discard this schedule in favor of the schedules with lower loop depths. This case highlights the need for a robust auto-scheduler to consider factors beyond loop depth.

\heading{Case \textcircled{4}: Neither the lowest loop depth, nor the highest loop depth schedule (Figure~\ref{fig:b-tensor-contraction-kernel}) is the best}
\label{experiment_4}
Setting loop bounds and sparsity as $I = 265$, $J = 1207$, $K = 479$, $L = 251$, $M = 234$, $N = 42$, and $\sparsity(B) \approx 0.0033$, Figure~\ref{fig:b-tensor-contraction-kernel} performs the fastest at $\Psi(b) = 513ms$, followed by $\Psi(d) = 1.14s$ for Figure~\ref{fig:c-tensor-contraction-kernel}, and $\Psi(a) = 1.66s$ for Figure~\ref{fig:a-tensor-contraction-kernel}.
In this scenario, auxiliary memories account for less than $12\%$ of the LLC. 
For these values, $\Phi(a) = 1.13 \times \Phi(b)$ and $\Phi(d) = 40.4 \times \Phi(b)$.
Execution times align with iteration complexities, and auxiliary memory usage is reasonably modest.
Unlike previous cases, where the best loop or memory depth proved to be the most efficient, this instance underscores the need for schedulers to consider multiple factors beyond loop and auxiliary memory depth when pruning the search space.

\subsection{Our approach: \system}
\label{sec:sparsetda}

The insights drawn from the motivating example and our approach to schedule selection can be summarized as follows.


\heading{Multi-Level Branched Loop Nests}
Nested loop computations with reduced loop depth (as in Case~\textcircled{1}) are crucial. 
However, existing scheduling languages lack support for multi-level branched loop nests. 
To address this, we extend the branched iteration graph (BIG)~\cite{taco,sparselnr} to accommodate recursive, multi-level branched iteration graphs with multi-dimensional temporary buffers. 
We also enhance the scheduling language to support recursive fusion by adapting TACO 's~\cite {taco} code generation strategies.

\heading{Time and Auxiliary Memory Complexities}
Both time and auxiliary memory complexities contribute to the schedule's execution time. 
An effective auto-scheduler needs to consider both aspects when selecting a schedule (as seen in Cases~\textcircled{2}--\textcircled{4}). 
If a schedule's auxiliary memory takes up a large portion of the last-level cache, it tends to perform worse than the alternatives (as observed in Case~\textcircled{2}).
To address this, we introduce an auto-scheduler that employs an SMT solver. 
The solver is guided by the constraints of sparse computations and reasons about the partial orders of time and auxiliary memory complexity. 
This approach effectively prunes the search space, leading to the selection of schedules that dominate others in both time and memory complexity.

\section{Design of the Transformation}\label{detailed_design}

Tensor contractions can be materialized using a simple linear loop nest where there would be a corresponding loop for each of the indices in the Einsum expression. 
This loop nest is represented as a linear iteration graph (LIG) as explained in Section~\ref{equivalence}, which is used for sparse code generation. 
However, this simple loop nest must respect the sparse tensor access constraints. For example, if a sparse tensor is in CSR format, the row should be accessed first. 
In this section, we describe an algorithm to recursively generate a branched iteration graph (BIG), the transformation required to convert a LIG into a multi-level BIG (\S~\ref{big}), and how this transformation can cover a plethora of possible loop nests for the computation (\S~\ref{completeness}).

\subsection{Linear Iteration Graph (LIG)~---Equivalence Class of Tensor Contractions}\label{equivalence}

Consider a tensor contraction:
\[O(\idx_{\mathsf{out}}) = \displaystyle\sum_{\idx_{\mathsf{contract}}} I_1(\idx_1)*\dots*I_i(\idx_i)*\dots*I_n(\idx_n).\]
Here, $I_1 \ldots I_n$ denote the input tensors; $O$ denotes the output tensor; $\idx_{contract}$ denotes the indices that need to be contracted from the tensor expression.
The example tensor contraction can be materialized in several ways, two of which are as follows (access indices are omitted for brevity):
\begin{figure}[h]
	\vspace{-1em}
    \centering
    \begin{adjustbox}{minipage=\linewidth,scale=1.0}
    \begin{subfigure}[h]{0.5\textwidth}
		\begin{lstlisting}[mathescape]
		$\loopC_1\ldots \loopC_i\ldots \loopC_j\ldots \loopC_n$:
		    $O$ += $I_1*...*I_l*...*I_m*...*I_e$
		\end{lstlisting}
		\vspace{-1em}
		\label{fig:lig-i-before-j}
    \end{subfigure}
    \hspace{5pt}
    \begin{subfigure}[h]{0.5\textwidth}
		\begin{lstlisting}[mathescape]
		$\loopC_1\ldots \loopC_j\ldots \loopC_i\ldots \loopC_n$:
		    $O$ += $I_1*...*I_m*...*I_l*...*I_e$
		\end{lstlisting}
		\vspace{-1em}
		\label{fig:lig-j-before-i}
    \end{subfigure}
    \end{adjustbox}
    \vspace{-1em}
\end{figure}

There are two main differences between the two materializations:
loops $i$ and $j$ are swapped, as well as tensors $I_l$ and $I_m$ in the expression.
Therefore, the orders of accessing elements of input tensors and storing elements of the output tensor differ.
But in general, any permutation of $loop_1$, $loop_2$, \dots, $loop_n$, and any permutation of $I_1$, \dots, $I_n$ yields the correct output tensor $O$, when we complete all the iterations.
This observation also holds when some of the tensors are sparse, although the index order must satisfy the sparse tensor access constraints.
Overall, materializations like the ones above belong to an \textit{equivalence class} because they produce the same output.

We define a \textbf{linear iteration graph (LIG)} as a loop nest with no two loops having the same depth from the root of the nest and an index order that respects all the sparse tensor access constraints.
Hence, we consider any permutation of the loops and input tensors that do not violate the sparse access constraints as a representative of an equivalence class since it produces the same result for a given tensor contraction.

\subsection{Multi-level Branched Iteration Graphs (BIG)}\label{big}

\begin{figure*}[t]
    \centering
    \begin{adjustbox}{minipage=\linewidth,scale=0.99}
    \begin{subfigure}[t]{0.108\textwidth}
      \centering
      \includegraphics[width=\textwidth]{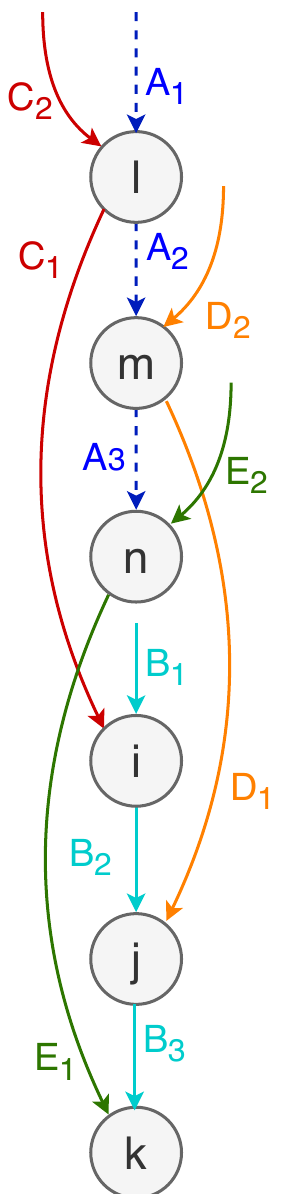}
      \caption{}
      \label{fig:tensor-contract-default}
    \end{subfigure}
    \hfill
    \begin{subfigure}[t]{0.162\textwidth}
      \centering
      \includegraphics[width=\linewidth]{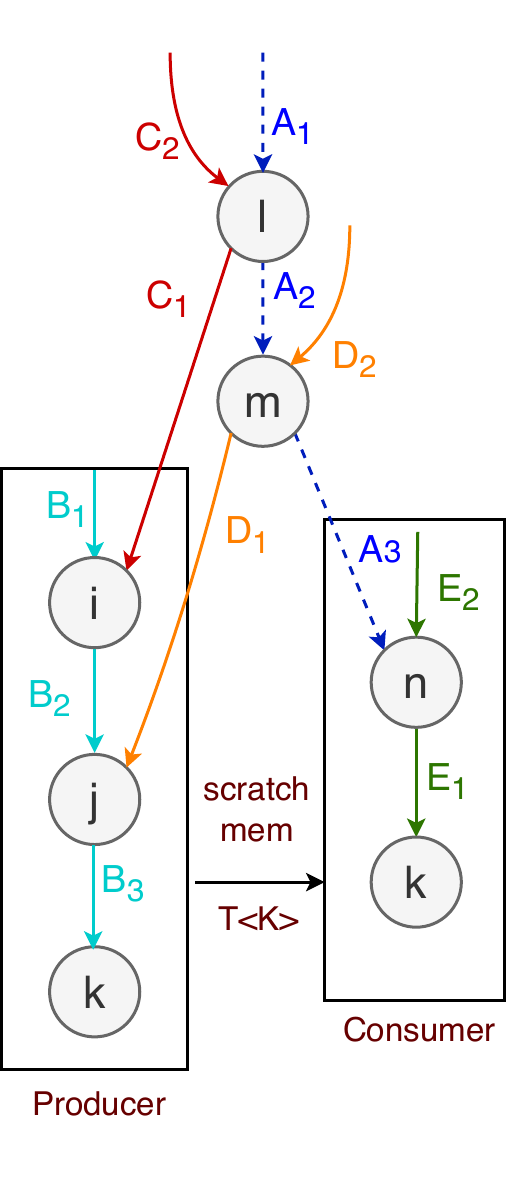}
      \caption{}
      \label{fig:tensor-contract-branched-k-mem}
    \end{subfigure}
    \hfill
    \begin{subfigure}[t]{0.162\textwidth}
      \centering
      \includegraphics[width=\linewidth]{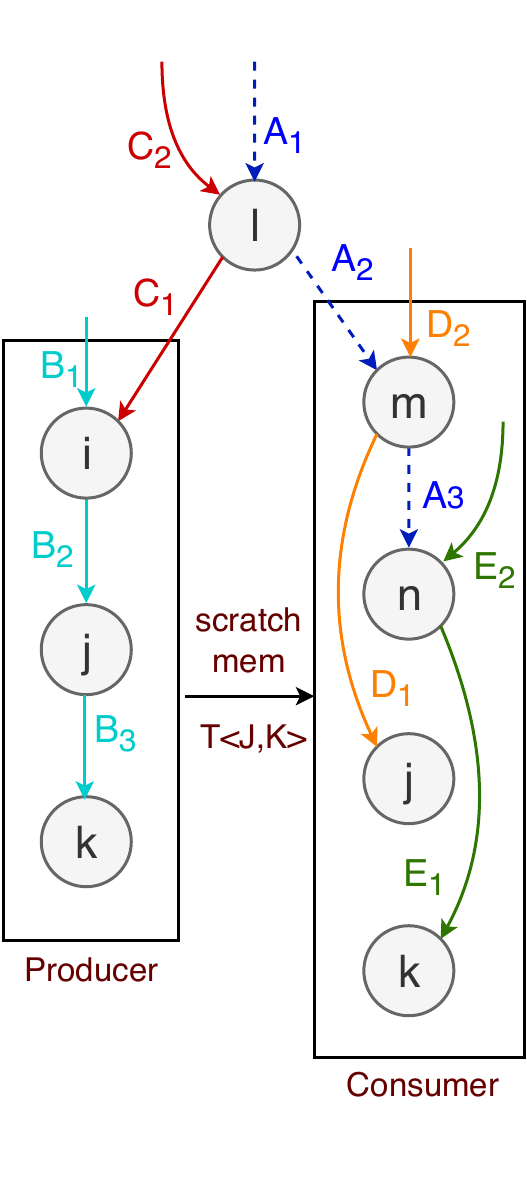}
      \caption{}
      \label{fig:tensor-contract-branched-jk-mem}
    \end{subfigure}
    \hfill
    \begin{subfigure}[t]{0.162\textwidth}
      \centering
      \includegraphics[width=\linewidth]{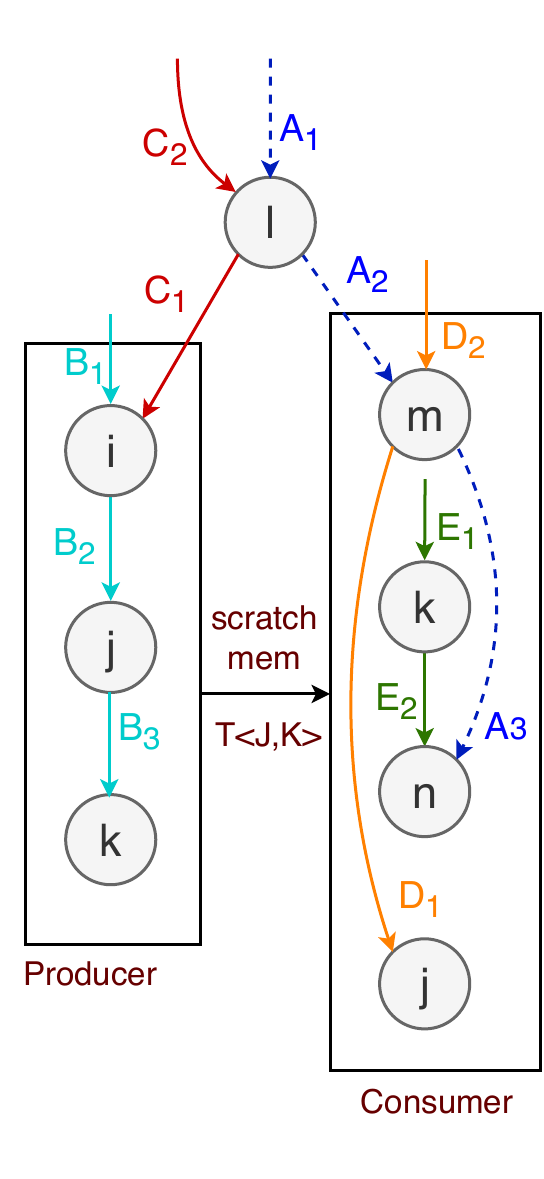}
      \caption{}
      \label{fig:tensor-contract-branched-jk-mem-reordered}
    \end{subfigure}
    \hfill
    \begin{subfigure}[t]{0.216\textwidth}
      \centering
      \includegraphics[width=\linewidth]{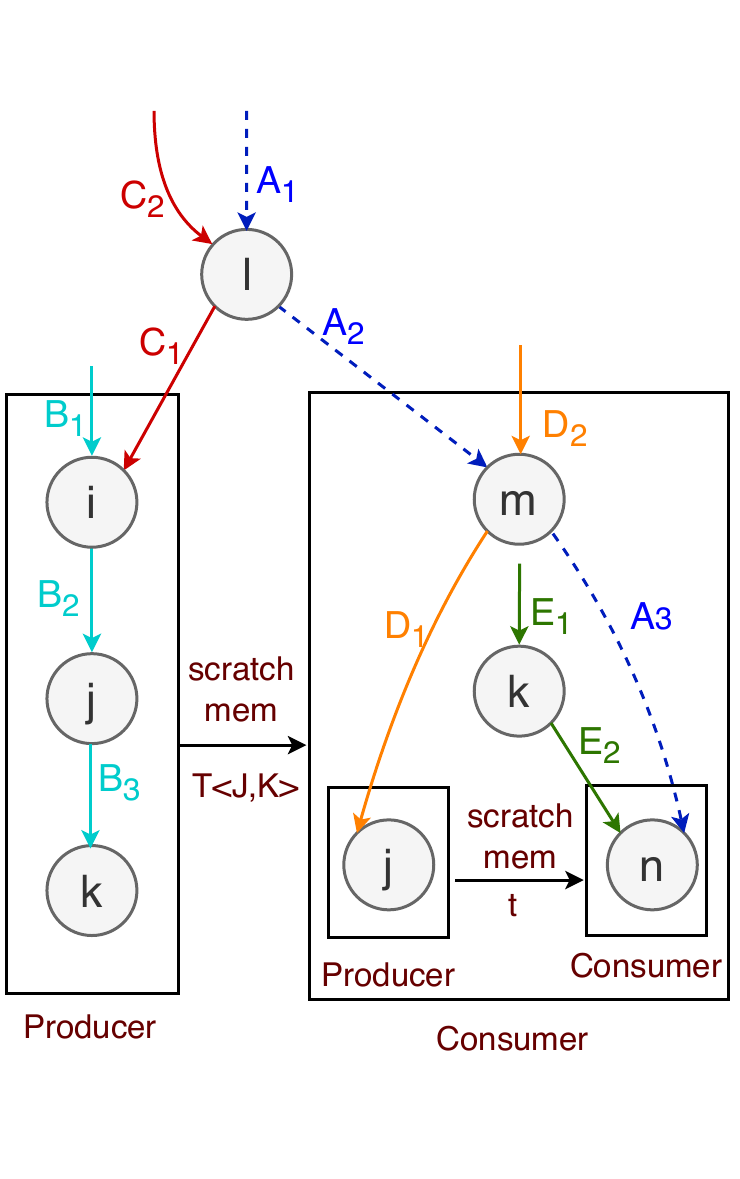}
      \caption{}
      \label{fig:tensor-contract-branched-jk-mem-final}
    \end{subfigure}
    \end{adjustbox}
    \caption{loopfuse transformation performed on $ A_{lmn} = \sum\nolimits_{ijk} {\sparse B_{ijk}} C_{il} D_{jm} E_{kn} $. (a) TACO default kernel, (b) Fused kernel with $K$ extra memory, (c) Fused kernel with $JK$ extra memory, (d) \ref{fig:tensor-contract-branched-jk-mem} with reordered consumer branch, and (e) Multi-level nesting after fusing inner branch of~\ref{fig:tensor-contract-branched-jk-mem-reordered}.}
    \label{fig:tensor-contract-all-graphs}
  \end{figure*}

\begin{figure}[t]
\vspace{-1.0em}
\begin{adjustbox}{minipage=1.01\columnwidth,scale=0.99}
\begin{algorithm}[H]
	\caption{Multi-level BIG Transformation}
    \label{algorithm:sparsefd}
    \small
    \begin{algorithmic}[1]
    \Require{Valid Iteration Graph $G^{complete}_{I}$}
    \Require{Path to Inner Iteration graph $path: Vector$}
    \Require{Split Position $i: Int$}
    \Require{Is Producer On the Left? $pol: bool$}
    \Ensure{Multi-level BIG $G'_{I}$}
    \vspace{0.5em}
    \State $G^{old}_{I} = GetInnerGraphUsingPath(G^{complete}_{I}, path)$ \label{lst:line:1}
    \State $comp = GetInnerComputation(G^{old}_{I})$ \Comment{computation: $A_{out} += A_{1}*A_{2}*..*A_{i}*...*A_{n}$} \label{lst:line:2}
    \State $expr_{producer} = (pol==True) : A_1*...*A_i\ ?\ A_{i+1}*...*A_n$ \label{lst:line:3}
    \State $expr_{consumer} = (pol==False) : A_{i+1}*...*A_n\ ?\ A_1*...*A_i$ \label{lst:line:4}
    \State $I_{temp} = GetIndices(Expr_{consumer})\cap(GetIndices(Expr_{producer})\cup GetIndices(A_{out}))$ \label{lst:line:5}
    \State $ProducerExpr :=  T'(I_{temp}) += expr_{prd}$ \label{lst:line:6}
    \State $ConsumerExpr := A_{out} += (pol==True): T'(I_{temp}) * expr_{consumer}\ ?\ expr_{consumer} * T'(I_{temp})$ \label{lst:line:7}
    \State $List_{I-Prd} = GetIndicesInOrder(ProducerExpr, G^{old}_{I})$ \label{lst:line:8}
    \State $List_{I-Con} = GetIndicesInOrder(ConsumerExpr, G^{old}_{I})$ \label{lst:line:9}
    \State \textbf{Define}: $I_{fusible} = \emptyset$ \label{lst:line:10}
    \ForEach {$i \in G^{old}_{I}$} 
        \If{$i \in List_{I-Prd} \textbf{ and } i \in List_{I-Con}$}
            \State $I_{fusible} = I_{fusible} \cup i$
        \Else \quad break;
        \EndIf
    \EndFor \label{lst:line:14}
    \State \textbf{Define}: $I_{shared} = \{List_{I-Prd} \cap List_{I-Con}\} \setminus I_{fusible}$ \label{lst:line:15}
    \State \textbf{Define}: $T(I_{shared})$ \label{lst:line:16}
    \State $ProducerExpr := T(I_{shared}) = expr_{producer}$ \label{lst:line:17}
    \State $ConsumerExpr := A_{out} = (pol==True): T(I_{temp}) * expr_{consumer}\ ?\ expr_{consumer} * T(I_{temp})$ \label{lst:line:18}
    \State $G^{new}_{I} = GraphRewrite(G^{old}_{I}, I_{fusible},ProducerExpr, ConsumerExpr)$ \label{lst:line:19}
    \State \Return{$GraphReplace(G^{complete}_{I}, G^{new}_{I}, G^{old}_{I})$} \label{lst:line:20}
    \end{algorithmic}
\end{algorithm}
\end{adjustbox}
\end{figure}

In this section, we describe the multi-level BIG transformation. 
We demonstrate how the algorithm works for the example tensor contraction from Section~\ref{overview} and by showing how the 
LIG in Figure~\ref{fig:tensor-contract-all-graphs} transforms 
into a BIG (\ref{fig:tensor-contract-default} $\rightarrow$ ~\ref{fig:tensor-contract-branched-k-mem}
and~\ref{fig:tensor-contract-default} $\rightarrow$~\ref{fig:tensor-contract-branched-jk-mem}
$\rightarrow$ ~\ref{fig:tensor-contract-branched-jk-mem-reordered} 
$\rightarrow$ ~\ref{fig:tensor-contract-branched-jk-mem-final}).

The tensor contraction $A(l,m,n) = {\sparse B(i,j,k)} * C(i,l) * D(j,m) * E(k,n)$
has a default iteration graph (Figure~\ref{fig:tensor-contract-default})
which implies generated code in Figure~\ref{fig:a-tensor-contraction-kernel}. The IR of this iteration graph is shown in the listing below:
\begin{lstlisting}[basicstyle=\small, gobble=3, tabsize=2, showtabs=false, showstringspaces=false,morekeywords={forall,where}]
    forall(l, forall(m, forall(n, forall(i, forall(j, forall(k, A(l,m,n) += B(i,j,k) * C(i,l) * D(j,m) * E(k,n)))))))
\end{lstlisting}

\heading{Transform Iteration Graph~\ref{fig:tensor-contract-default} $\rightarrow$ ~\ref{fig:tensor-contract-branched-k-mem}}
Let us split the computation into two parts, the producer and the consumer.
The first one, $T(\idx_{\mathsf{temp}}) = \sum {\sparse B(i,j,k)} * C(i,l) * D(j,m)$, produces the intermediate temporary tensor $T$ with indices $\idx_{\mathsf{temp}}$, which is consumed in the second one, $A(l,m,n) = T(\idx_{\mathsf{temp}}) * E(k,n)$, to generate the output $A$.
Here, $\idx_{\mathsf{temp}}=\{l,m,k\}$  is obtained by evaluating:
\[\Indices({\sparse B(i,j,k)} * C(i,l) * D(j,m)) \cap (\Indices(A(l,m,n)) \cup \Indices(E(k,n))).\]
Hence, the auxiliary memory required is $O(LMK)$.

The split computations from the original one have iteration graphs with the following index orders: $\encircle{l}\veryshortarrow\encircle{m}\veryshortarrow\encircle{i}\veryshortarrow\encircle{j}\veryshortarrow\encircle{k}$ for the producer and $\encircle{l}\veryshortarrow\encircle{m}\veryshortarrow\encircle{n}\veryshortarrow\hspace{-2pt}\encircle{k}$ for the consumer.
These ones preserve the index order in the original iteration graph $\encircle{l}\veryshortarrow\encircle{m}\veryshortarrow\encircle{n}\veryshortarrow\encircle{i}\veryshortarrow\encircle{j}\veryshortarrow\encircle{k}$ (Figure~\ref{fig:tensor-contract-default}).
As both iteration graphs share the same indices $l$ and $m$ at the beginning, they can be fused into the BIG in Figure~\ref{fig:tensor-contract-branched-k-mem}, facilitating code generation in Figure~\ref{fig:b-tensor-contraction-kernel}.
Since the unfused sections of the producer and consumer graphs include $i,j,k$ and $n,k$ respectively, the fused iteration graph would require extra memory of $K$, obtained from $\{i,j,k\} \cap \{n,k\}$, to pass the intermediate results between the two computations. 
Notably, only an auxiliary memory of size $K$ is required after fusion, compared to the one with size $LMK$ before fusion. 
 

\heading{Transform Iteration Graph~\ref{fig:tensor-contract-default} $\rightarrow$ ~\ref{fig:tensor-contract-branched-jk-mem}}
The original tensor contraction can be split into two computations in a different way. 
For example, $T(\idx_{\mathsf{temp}}) = {\sparse B(i,j,k)} * C(i,l)$ and $A(l,m,n) = T(\idx_{\mathsf{temp}}) * D(j,m) * E(k,n)$ where $\idx_{\mathsf{temp}}=\{l,j,k\}$.
This would result in the producer and consumer iteration graphs $\encircle{l}\veryshortarrow\encircle{i}\veryshortarrow\encircle{j}\veryshortarrow\encircle{k}$ and $\encircle{l}\veryshortarrow\encircle{m}\veryshortarrow\encircle{n}\veryshortarrow\encircle{j}\veryshortarrow\encircle{k}$, respectively. 
Since they have a common index $l$ at the beginning of the iteration graph, they can be fused to generate the BIG in Figure~\ref{fig:tensor-contract-branched-jk-mem}. 
Fusing the iteration graph reduces the auxiliary memory requirement to $JK$ as opposed to the $LJK$ before. 
After fusion, the IR is shown in the listing below (notice the addition of the temporary $t1$ and the \code{where} clause to combine the producer and consumer computations):
\begin{lstlisting}[basicstyle=\small, gobble=3, tabsize=2, showtabs=false, showstringspaces=false,morekeywords={forall,where},escapechar=\%]
    forall(l, %\color{red}where%(
        forall(m, forall(n, forall(j, forall(k, A(l,m,n) += t1(j,k) * D(j,m) * E(k,n))))), 
        forall(i, forall(j, forall(k, t1(j,k) += B(i,j,k) * C(i,l))))))
\end{lstlisting} 

\heading{Transform Iteration Graph~\ref{fig:tensor-contract-branched-jk-mem} $\rightarrow$ ~\ref{fig:tensor-contract-branched-jk-mem-reordered}} We notice that, after fusion, each of the unfused sections of the producer and consumer iteration graphs can be treated as separate iteration graphs by keeping all the fused indices fixed in the computation. 
For example, take the consumer computation, $A(l,m,n) = T(l,j,k) * D(j,m) * E(k,n)$. 
We can rewrite the computation as $A(\_,m,n) += T(\_,j,k) * D(j,m) * E(k,n)$ by fixing $l$, with corresponding iteration graph $\encircle{m}\veryshortarrow\encircle{n}\veryshortarrow\encircle{j}\veryshortarrow\encircle{k}$.
This can be split into two computations: $T'(m,k) += T(\_,j,k) * D(j,m)$ and $A(\_,m,n) += T'(m,k) * E(k,n)$, with corresponding iteration graphs $\encircle{m}\veryshortarrow\encircle{j}\veryshortarrow\encircle{k}$ and $\encircle{m}\veryshortarrow\encircle{n}\veryshortarrow\encircle{k}$, respectively. 
Both of these iteration graphs have the same first index \encircle{m} that can be fused. 
The iteration graph in this configuration (not shown in Figure~\ref{fig:tensor-contract-all-graphs}) would be able to generate the code in Figure~\ref{fig:d-tensor-contraction-kernel}. 
This configuration would require auxiliary memory of size $K$ because the unfused part of each iteration graph shares the common index $\{k\} = \{j,k\} \cap \{n,k\}$.
The iteration graph in Figure~\ref{fig:tensor-contract-branched-jk-mem} can be transformed to the iteration graph in Figure~\ref{fig:tensor-contract-branched-jk-mem-reordered} with the inner consumer index order $\encircle{m}\veryshortarrow\encircle{k}\veryshortarrow\encircle{n}\veryshortarrow\encircle{j}$, by reordering the consumer part of Figure~\ref{fig:tensor-contract-branched-jk-mem}.

\heading{Transform Iteration Graph~\ref{fig:tensor-contract-branched-jk-mem-reordered} $\rightarrow$ ~\ref{fig:tensor-contract-branched-jk-mem-final}} Splitting the consumer computation as described previously yields the producer and consumer iteration graphs $\encircle{m}\veryshortarrow\encircle{k}\veryshortarrow\encircle{j}$ and $\encircle{m}\veryshortarrow\encircle{k}\veryshortarrow\encircle{n}$, respectively, which can be fused to generate the multi-level BIG in Figure~\ref{fig:tensor-contract-branched-jk-mem-final}.
Since the unfused sections of the producer and consumer graphs do not share any common indices, it only requires a scalar auxiliary memory to pass the intermediate results between the producer and consumer.
The final IR is shown in the listing below (notice the use of two temporaries $t1$ and $t2$, and the nested combination of \code{where} clauses):
\begin{lstlisting}[basicstyle=\small, gobble=3, tabsize=2, showtabs=false, showstringspaces=false,morekeywords={forall,where},escapechar=\%]
    forall(l, %\color{red}where%(  
        forall(m, forall(k, %\color{red}where%(  
            forall(n, A(l,m,n) += t2 * E(k,n)), 
            forall(j, t2 += t1(j,k) * D(j,m))))), 
        forall(i, forall(j, forall(k, t1(j,k) += B(i,j,k) * C(i,l))))))
\end{lstlisting} 

\subsection{LIG to BIG Transformation Algorithm}\label{lig2big}

Algorithm~\ref{algorithm:sparsefd} shows the pseudo-code for the transformation described previously in Section~\ref{big}.
This algorithm takes several inputs: the original iteration graph (LIG or BIG), the $path$ to an inner producer/consumer section, the position to split ($Split\ Position:\ i$), the input tensors in the contraction, and a boolean flag to indicate whether the producer expression is on the left or the right after the split.
Input variable $path$ is used in line~\ref{lst:line:1} to access the inner producer/consumer graph sections, which helps to apply the transformation {\em recursively} to the inner linear graph sections. 

The split operation occurs in lines~\ref{lst:line:2}--\ref{lst:line:4}. 
For example, given the expression $A=B*C*D$ and $i=2$, the splits are $T=B*C$ and $A=T*D$. 
If $i=1$, then the splits are $T=B$ and $A=T*C*D$. 
The algorithm initially deduces the indices of the temporary resulting from the split (line~\ref{lst:line:5}) using the equation $Indices(Consumer)\cap(Indices(Producer)\cup Indices(Output))$. 
This equation calculates the indices in the producer that also appear in either the consumer or the output. 
The algorithm generates corresponding split expressions in lines~\ref{lst:line:6}--\ref{lst:line:7}. 
Subsequently, the producer and consumer graphs are computed in lines~\ref{lst:line:8}--\ref{lst:line:9}, preserving the index order of the original graph. 
Then, the algorithm determines the fusible outer loops (lines~\ref{lst:line:10}--\ref{lst:line:14}) and shared indices (lines~\ref{lst:line:15}--\ref{lst:line:16}). 
Finally, the algorithm produces the expressions for the producer and consumer in the fused iteration graph in lines~\ref{lst:line:17}--\ref{lst:line:18}, and the original iteration graph is replaced with the fused iteration graph in lines~\ref{lst:line:19}--\ref{lst:line:20}. 
One step of the transformation is linear time with respect to the number of indices in the iteration graph.


\subsection{Scheduling Language}\label{schedulelang}

Figure~\ref{fig:tensor-contract-transformations} shows the implementation of the transformations described in Section~\ref{big} using the scheduling language.
The schedule description in Figure~\ref{fig:transformation-2} can be used to transform the iteration graph in Figure~\ref{fig:tensor-contract-default} to the one in Figure~\ref{fig:tensor-contract-branched-k-mem}.
The schedule description in Figure~\ref{fig:transformation-4} can be used to transform the original iteration graph in Figure~\ref{fig:tensor-contract-default} to the one in Figure~\ref{fig:tensor-contract-branched-jk-mem-final} by doing multiple transformations.

\heading{\texttt{loopfuse}} scheduling directive splits and fuses LIGs. It takes three parameters:
\begin{description}
\item[\texttt{path}]
identifies linear graph sections, consumer or producer sections, of a BIG.
The \verb|path| parameter must direct to a linear graph section for the transformation to be applied.
Here, $\{\}$ means accessing the root of an iteration graph, $\{0\}$ means accessing the producer section, and $\{1\}$ means accessing the consumer section. 
If the BIG has multiple levels, $\{0,1\}$ would access the producer of the first level and then the consumer of the second level.

\item[\texttt{loc}]
specifies the split position in the inner computation.
For example, if the inner computation is $A=B*C*D$ and \verb|loc=2|, then the split is $T=B*C$ and $A=T*D$, and if \verb|loc=1|, then the split is $T=B$ and $A=T*C*D$.

\item[\texttt{pol}]
designates the first or second half of the contraction as the producer.
If \verb|pol=True|, then the producer is on the left, and the consumer is on the right, and if \verb|pol=False|, then vice versa. 
For example, if the expression is $A=B*C*D$, \verb|loc=2| and \verb|pol=True|, then the split is $T=B*C$ and $A=T*D$, and if \verb|pol=False|, then the split is $T=C*D$ and $A=B*T$. 
\end{description}

\heading{\texttt{reorder}}
scheduling directive reorders indices of a linear graph section. It takes two parameters:
\begin{description}
\item[\texttt{path}]
identifies an inner linear graph section.
\item[\texttt{order}]
specifies the new order of the indices in the linear graph section.
\end{description}

The \verb|reorder| and \verb|loopfuse| directives can be used together to obtain the desired multi-level BIG. 
These two directives can be used in conjunction to generate all possible loop trees for a given tensor contraction.

\begin{figure}[t]
  \centering
  \begin{adjustbox}{minipage=\linewidth,scale=1.0}
    \begin{subfigure}[t]{0.45\textwidth}
      \begin{lstlisting}[basicstyle=\small, gobble=8, tabsize=2, showstringspaces=false, escapechar=\%]
        A(l,m,n) = B(i,j,k)*C(i,l)*D(j,m)*E(k,n);
        // Index stmt of %\ref{fig:tensor-contract-default}%
        IndexStmt stmt = A.getAssignment().concretize();
        // Apply transformation
        stmt = stmt // %\ref{fig:tensor-contract-default}% -> %\ref{fig:tensor-contract-branched-k-mem}%
          .loopfuse(loc = 2, pol = True, path = {});
      \end{lstlisting}
      \vspace{0.8em}
      \subcaption{\ref{fig:tensor-contract-default} $\rightarrow$ \ref{fig:tensor-contract-branched-k-mem}}
      \label{fig:transformation-2}
    \end{subfigure}
    \begin{subfigure}[t]{0.55\textwidth}
      \begin{lstlisting}[basicstyle=\small, gobble=8, tabsize=2, showstringspaces=false, escapechar=\%]
        A(l,m,n) = B(i,j,k)*C(i,l)*D(j,m)*E(k,n);
        // Index stmt of %\ref{fig:tensor-contract-default}%
        IndexStmt stmt = A.getAssignment().concretize();
        // Apply transformation
        stmt = stmt
          .loopfuse(loc = 3, pol = True, path = {}) // %\ref{fig:tensor-contract-default}% -> %\ref{fig:tensor-contract-branched-jk-mem}%
          .reorder(order = {m, k, n, j}, path = {1}) // %\ref{fig:tensor-contract-branched-jk-mem}% -> %\ref{fig:tensor-contract-branched-jk-mem-reordered}%
          .loopfuse(loc = 2, pol = True, path = {1}); // %\ref{fig:tensor-contract-branched-jk-mem-reordered}% -> %\ref{fig:tensor-contract-branched-jk-mem-final}%
      \end{lstlisting}
      \vspace{-0.5em}
      \subcaption{\ref{fig:tensor-contract-default} $\rightarrow$ \ref{fig:tensor-contract-branched-jk-mem} $\rightarrow$ \ref{fig:tensor-contract-branched-jk-mem-reordered} $\rightarrow$ \ref{fig:tensor-contract-branched-jk-mem-final}}
      \label{fig:transformation-4}
    \end{subfigure}
  \end{adjustbox}
  \vspace{-1em}
  \caption{Transformation on the loop contraction}
  \vspace{-1em}
  \label{fig:tensor-contract-transformations}
\end{figure}

\subsection{Completeness of the algorithm}\label{completeness}

This section provides a proof sketch for the completeness of Algorithm~\ref{algorithm:sparsefd}: we argue that the algorithm can generate all possible loop structures for a given tensor contraction using the \verb|loopfuse| and \verb|reorder| scheduling directives.

\heading{Constraints}
Two essential constraints ensure the validity of BIG constructed by Algorithm~\ref{algorithm:sparsefd} and the equivalence of the BIG to the initial LIG.
First, the BIG must not violate any sparse tensor access constraints present in the initial computation.
For instance, in all the iteration graphs in Figure~\ref{fig:tensor-contract-all-graphs}, the iteration order $B_1\veryshortarrow B_2\veryshortarrow B_3$ (\ie $i\veryshortarrow j\veryshortarrow k$), is consistently maintained when contracting the sparse tensor ${\sparse B}$ with other tensors.
Second, a permutation of indices in producer and consumer loops is required to establish identical orders of shared indices (\ie indices in the temporary tensor).

The second constraint can be understood as follows. Let $i$ and $j$ be the indices present in the temporary tensor.
Let $P$ and $C$ be the sets of all valid (in the sense of the first constraint) permutations of indices in the loops of producer and consumer, respectively.
Let the focus be on $i\veryshortarrow\cdots\veryshortarrow j$ and $j\veryshortarrow\cdots\veryshortarrow i$ in $P$ and $C$ sets.
Let {\em condition1} be $i\veryshortarrow\cdots\veryshortarrow j$ in $P$ and $i\veryshortarrow\cdots\veryshortarrow j$ in $C$, and {\em condition2} be $j\veryshortarrow\cdots\veryshortarrow i$ in $P$ and $j\veryshortarrow\cdots\veryshortarrow i$ in $C$. 
If either {\em condition1} or {\em condition2} is satisfied, then we say that the BIG is valid. 
If neither {\em condition1} nor {\em condition2} is satisfied, then we say the second constraint is violated and the BIG is invalid.
The temporary tensors introduced by the Algorithm~\ref{algorithm:sparsefd} are dense. 
Hence, they do not impose extra constraints, and establishing identical orders of shared indices is not impeded by the temporaries.

\begin{figure}[t]
  \begin{adjustbox}{minipage=\linewidth,scale=0.99}
  \centering
      \begin{subfigure}[h]{0.31\textwidth}
  \begin{lstlisting}[mathescape]
  $loops\ P$:
    $T(S)+=Comp_P$
  $loops\ C$:
    $A(..)+=T(S)*Comp_{C}$
  \end{lstlisting}
        \vspace{-0.5em}
        \caption[LIG with i loop before j loop]{}
        \label{fig:a-big-to-lig}
        \vspace{-15pt}
      \end{subfigure}
      \begin{subfigure}[h]{0.31\textwidth}
  \begin{lstlisting}[mathescape]
  $T(S) = \sum_{P\setminus S} Comp_P$
  $loops\ C$:
    $A(..)+=T(S)*Comp_{C}$
  \end{lstlisting}
        \caption[LIG with j loop before i loop]{}
        \label{fig:b-big-to-lig}
        \vspace{-15pt}
      \end{subfigure}
      \begin{subfigure}[h]{0.31\textwidth}
  \begin{lstlisting}[mathescape]
  $loops\ C$:
    $A(..)+=$ 
      $(\sum_{P\setminus S} Comp_P)*Comp_{C}$
  \end{lstlisting}
        \caption[LIG with j loop before i loop]{}
        \label{fig:c-big-to-lig}
        \vspace{-15pt}
      \end{subfigure}
      \vskip\baselineskip
      \begin{subfigure}[h]{0.31\textwidth}
  \begin{lstlisting}[mathescape]
  $loops\ C$:
    $A(..)+=$ 
      $(\sum_{P\setminus S} Comp_P*Comp_{C})$
  \end{lstlisting}
        \vspace{-0.5em}
        \caption[LIG with j loop before i loop]{}
        \label{fig:d-big-to-lig}
        \vspace{-1em}
      \end{subfigure}
      \hspace{5pt}
      \begin{subfigure}[h]{0.31\textwidth}
  \begin{lstlisting}[mathescape]
  $loops\ C \cup (P\setminus S)$:
    $A(..)+=Comp_P*Comp_{C}$
  \end{lstlisting}
        \vspace{-0.2em}
        \caption[completed BIG to LIG transformation]{}
        \label{fig:e-big-to-lig}
        \vspace{-1em}
      \end{subfigure}
    \end{adjustbox}
    \caption{Moving producer computation to the consumer to obtain a LIG.}
    \vspace{-1em}
    \label{fig:big-to-lig}
\end{figure}

\heading{Equivalence of BIG and LIG}
In the context of reasoning about LIGs and BIGs, having a transformation from BIG to LIG, complementary to the one performed by Algorithm~\ref{algorithm:sparsefd}, proves beneficial.
This transformation involves examining the innermost subtree (with no nested subtrees inside of it), as illustrated in Figure~\ref{fig:big-to-lig}.
Denoting shared indices between the producer and consumer as $S$, producer loops as $P$, and consumer loops as $C$, the transformation requires an order of $S$ complying with sparse tensor access constraints in both $P$ and $C$, adhering to the second constraint.
By keeping the outer loops constant in inner branch computations (details omitted for simplicity), the BIG illustrated in Figure~\ref{fig:a-big-to-lig} is realized.
The contraction for $T$ is depicted in Figure~\ref{fig:b-big-to-lig}, where $P\setminus S$ denotes contracted indices.
Substituting $T(S)$ inside the consumer (Figure~\ref{fig:c-big-to-lig}), the consumer computation ($Comp_c$) is moved inside the summation operation, ensuring none of its indices contain the ones in $P\setminus S$ (Figure~\ref{fig:d-big-to-lig}).
Further, contracting indices are reintegrated into consumer graph loops (Figure~\ref{fig:e-big-to-lig}), satisfying all constraints and resulting in a LIG. This recursive process applies to multi-level BIGs, yielding a LIG equivalent to the initial BIG.

\heading{Fineteness of the space}
A conservative upper bound on the number of BIGs can be
established by considering the number of input tensors in the tensor contraction ($n$), 
and the number of indices in the tensor contraction ($m$). The input tensors can be 
permuted in $n!$ ways. A BIG can be built by 
recursively splitting the computation into producer and consumer sections. Since the number
of input tensors are $n$, the number of binary trees that can be built is bounded by 
$2^{n}$. The indices can be permuted in $m!$ ways. Since there are $2^{n}$ splits, indices can be 
permuted at each split giving $m!^{2^{n}}$ permutations. At each split operation, there is a choice to fuse the indices or not. Since 
the number of indices is $m$, the number of ways to fuse the indices is bounded by $m+1$ for
each of those binary trees. 
Therefore, the total number of BIGs is bounded by
$n! \times 2^{n} \times m!^{2^{n}} \times (m+1)$.

\heading{Completeness}
Consider the different schedules for a given tensor contraction as points in a space.
If you can reach a schedule from another, then they are connected in this space. 
We established that a BIG can be linearized. Therefore, every BIG is connected to a LIG.
As outlined in Section~\ref{equivalence}, LIG schedules are equivalent, and we end up connecting all the points. 
Focusing on the linearization procedure for a BIG, the movement of $T$ from producer to consumer involves reordering the loops of producer and consumer such that after reordering the loops, the shared indices have the same relative ordering. 
$T$ consists of some input tensors in the original tensor contraction. 
In other words, input tensors in the producer computation are some combination of the input tensors. 
This combination can be obtained by permuting the input tensors in the original expression and splitting from a specific position. 
Since each valid BIG can be transformed to a LIG, it is possible to traverse in the direction of LIG to BIG by using the transformation in Sections~\ref{big}--\ref{schedulelang}.
Therefore, by (1) permuting all schedules in our equivalence class, (2) applying the transformation in Section~\ref{big} to obtain BIGs, and (3) recursively applying (1) and (2) on inner producer and consumer sections, we can generate all possible iteration graphs (loop structures of schedules) for that computation.

\section{Auto-Scheduler}\label{autoscheduler}


We build an auto-scheduler that, given a tensor contraction, explores the schedule space and chooses a memory- and time-efficient schedule.
The main function of the scheduler is pruning the schedule space.
The scheduler decides on what schedules to prune by creating a Pareto frontier of schedules using partially ordered sets of time and memory complexity.

\begin{figure}[t]
	\centering
    \includegraphics[width=1.0\linewidth]{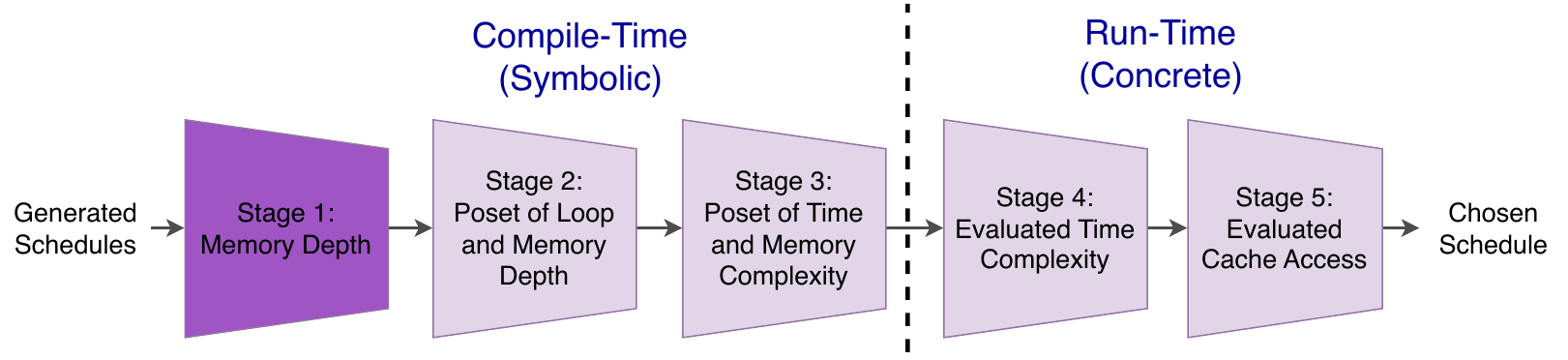}
    \caption{Pruning Stages of the auto-scheduler. Stages 2 to 5 compare a schedule against others for pruning. Stage 1 uses an absolute measure to filter schedules without comparing them.}
    \label{fig:pruning_stages}
\end{figure}

The complete pruning pipeline is shown in Figure~\ref{fig:pruning_stages}.
The pipeline starts by generating schedules in the search space (\S~\ref{schedule_generation}).
The following stages are divided into two parts.
The first three stages are executed during compile-time with symbolic expressions (\S~\ref{symbolic_stages}), and the last two stages are executed with concrete expressions at run-time (\S~\ref{concrete_stages}).

\subsection{Schedule Generation}\label{schedule_generation}

For a given expression and an LIG, we generate all the loop index orders that conform to the sparse tensor access constraints.
Then, we split the tensor contraction at different positions for all those index orders. 
Once tensor contraction is split, we infer the temporary indices and call the same function recursively for both the consumer and producer sub-computations.
After those sub-computations return the consumer schedules $C_{\Schedules}$ and producer schedules $P_{\Schedules}$, we combine those two schedule spaces as $P_{\Schedules} \times C_{\Schedules}$ to create super schedules that completely describe the initial computation.
If the producer and consumer sections can be fused as explained in Section~\ref{big}, we merge the sub-computations to create fused schedules, which we add to the list of schedules.

Consider the previous example (\S~\ref{overview}-\ref{detailed_design}). 
First, we create different permutations of input tensors. Since tensor contractions are commutative in Einsum notation, $A(l,m,n) = {\sparse B(i,j,k)} * C(i,l) * D(j,m) * E(k,n)$ is equivalent to $A(l,m,n) = {\sparse B(i,j,k)} * E(k,n) * C(i,l) * D(j,m)$. 
For each of these permutations, we create permutations of indices that conform to sparse tensor access constraints. 
These two steps combined create the complete set of LIGs (\S~\ref{equivalence}). 
For each LIG, we split the input tensors at different positions to generate producers and consumers. 

There are two ways in which we can split the input tensors. 
$A = B*C*D*E$ can be split as (a) $T = B*C$, and $A = T*D*E$: the result of the producer is directly used in consumer, and (b) $T1 = B*C$, $T2 = D*E$, and $A = T1*T2$: the consumer expects results of two producers. 
This procedure can be repeated ({\em recursive application of the algorithm}) on the producer and consumer sub-computations. 
Out of these two ways, the first one is more interesting because it opens up avenues for loop fusion. 
If the producer and consumer graphs contain the same indices, then we can fuse them. 
Consider $A(l,m,n) = {\sparse B(i,j,k)} * C(i,l) * D(j,m) * E(k,n)$ with index order $l \veryshortarrow m \veryshortarrow n \veryshortarrow i \veryshortarrow j \veryshortarrow k$, split between $C$, and $D$. 
The fusion of them would result in $l \veryshortarrow \langle T(j,k); producer: i \veryshortarrow j \veryshortarrow k: T(j,k) += B(i,j,k) * C(i,l), consumer: m \veryshortarrow n \veryshortarrow j \veryshortarrow k: A(l,m,n) += T(j,k) * D(i,l) * E(k,n)\rangle$ (\S~\ref{big}). 
After the fusion operation, we remove the fused indices in the inner computation, for example, $A(\_, m, n) += T(j, k) * D(i, \_) * E(k, n)$, to recursively call the schedule generation procedure, and combine the results with the outer loops.

\subsection{Symbolic Stages}\label{symbolic_stages}

\heading{Memory-Depth-Based Pruning}\label{memory_dim_based_pruning}
The first pruning stage analyzes the dimensions of the temporary tensors used in the schedule.
For example, Schedule~\ref{fig:a-tensor-contraction-kernel} does not use any temporaries, and hence, its memory depth is $0$.
In Schedule~\ref{fig:b-tensor-contraction-kernel}, the temporary $T\langle K \rangle$ has the memory depth of $1$, whereas Schedules~\ref{fig:c-tensor-contraction-kernel} and \ref{fig:d-tensor-contraction-kernel} have the memory depth of $2$ since they employ 2D temporary tensors.
If we split the computation as before to $T_{jkl} = B_{ijk} * C_{il}$ and $A_{lmn} = T_{jkl}*D_{jm}*E_{kn}$ but do not fuse (see Figure~\ref{fig:tensor-contract-branched-k-mem} for the fused version), the memory depth would be $3$.
The memory depth $3$ is too high for most realistic scenarios, and hence,
we use it as a threshold to prune schedules like that.

The memory-depth-based heuristic goes first in the pruning pipeline because it usually discards many unrealistic schedules.
At the end of this stage, we compute the symbolic iteration time and auxiliary memory complexity for each of the schedules. 
Then, we allocate the schedules into groups using the symbolic (time, auxiliary memory) complexity tuples.

We note that this stage can be replaced by a more sophisticated memory-volume-based pruning mechanism that uses an SMT solver. With such an approach, the user can provide a heuristic upper bound value on the total auxiliary memory use.
This type of pruning would guarantee that the schedules with higher depth but lesser auxiliary memory are not pruned and the schedules with lesser depth but higher auxiliary memory are pruned. 
However, this would require further exploration on how to select the upper bound value which often depends on the execution environment (\ie machine parameters, such as cache size, etc.).
Moreover, a memory-based pruning stage capable of reasoning about the actual volume of auxiliary memory
will require the complete information about the loop bounds.
Knowing either piece of information---machine parameters or loop bounds---blurs the boundary between compile-time and run-time stages and goes beyond our approach.
We leave this idea for future work.

\heading{Loop- and Memory-Depth-Based Pruning}\label{time_and_memory_depth_based_pruning}
In Section~\ref{overview}, we explained that using only the loop depth could prune potentially useful schedules.
Therefore, at this stage, we consider both the loop depth and memory depth for pruning to create a Pareto frontier of schedules for the next stage.
We use a poset to remove the schedules that are worse in terms of both loop depth and memory depth.
The poset-based pruning mechanism ensures that pruned schedules contain linear loop nest schedules (with no branches), including the default TACO schedule, as long as there are no other schedules with a scalar auxiliary memory and lesser time complexity.
This guarantees that we end up with a superior schedule at the end when compared to the default schedule.
The memory depth heuristic we used in Stage~1 (Section~\ref{memory_dim_based_pruning}) ensures that we do not prune schedules that are likely to have lesser loop depth than the fused simple linear loop nest schedules.

The poset-based pruning mechanism can be formally written as follows.
This stage removes a schedule $s$ from the set of schedules $S$ (received from Stage 1) if
there exists $c \in S$ such that
\[(L(s) > L(c) \land M(s) >= M(c)) \lor (L(s) >= L(c) \land M(s) > M(c))\] 
where $L(s)$ and $M(s)$ are the loop and memory depths of the schedule $s$, respectively.

This type of pruning ensures that we do not remove schedules that are likely to be better in the Pareto frontier of schedules.
We allocate each schedule to a different (time, memory) bucket at the end of this stage.
In other words, the schedules in a bucket have the same iteration time and memory complexity but differ in the order of loops.

\begin{figure*}[t]
\begin{lstlisting}[language = Python, basicstyle=\small, gobble=0, tabsize=2, showstringspaces=false, escapechar=\%, commentstyle=\color{darkgray}, caption={Formulation of the Pareto Frontier using the SMT solver.}, captionpos=b, label={lst:z3_pruning}]
# Definition of Range (R), Inferred (I), and User-Defined (U) constraints
R = (i >= 1e3) %$\wedge$% (i <= 1e6) %$\wedge$% (sparsity_B >= 0.001) %$\wedge$% (sparsity_B <= 0.01) %$\wedge$% (nnz_B <= 10e4) %$\wedge$% ... %\label{line:rc}%
I = (j_pos < j) %$\wedge$% (k_pos < k) %$\wedge$% (nnz_B < i*j*k) %$\wedge$% ... %\label{line:ic}%
U = (10*j < k) %$\wedge$% ... %\label{line:uc}%

# Obtain time and auxiliary memory complexities of schedules in Figure%~\ref{fig:tensor-contract-default}% (s1) and Figure%~\ref{fig:tensor-contract-branched-jk-mem-final}% (s2)
t1 = l*m*n*nnz_B; m1 = 0
t2 = l*nnz_B + l*m*k*j + l*m*k*n; m2 = j*k %\label{line:comp}%

# Formulation of the Pareto frontier:
c1 = (t1 >= t2); c2 = (m1 > m2); c3 = (t1 > t2); c4 = (m1 >= m2) %\label{line:cond}%
cond1 = R %$\wedge$% I %$\wedge$% U %$\wedge$% ((c1 %$\wedge$% c2) %$\vee$% (c3 %$\wedge$% c4)); 
cond2 = R %$\wedge$% I %$\wedge$% U %$\wedge$%  ((%$\neg$%c1 %$\wedge$% %$\neg$%c2) %$\vee$% (%$\neg$%c3 %$\wedge$% %$\neg$%c4)) %\label{line:cond-final}%

if (cond1 is SAT and cond2 is UNSAT): # s1 is dominated by s2. Remove s1 from the Pareto frontier. %\label{line:sat}%
else if (cond1 is UNSAT and cond2 is SAT): # s2 is dominated by s1. Remove s2 from the Pareto frontier. %\label{line:unsat}%
else: # s1 and s2 are incomparable. Keep both s1 and s2 in the Pareto frontier. %\label{line:nocomare}%
\end{lstlisting}
\end{figure*}

\heading{SMT-Solver-Based Pruning}\label{z3_based_pruning}
In some cases, the user (\eg performance engineer) may know during the compile-time some information about the loop bounds or sparsities of the tensors used in the computation.
For example, if it is a graph neural network computation, the user may know that the feature size of the nodes is in the range of $[16, 256]$, or the graph size is in the order of $10M$ and the sparsity of the graph is in the range of $[0.001, 0.01]$.
Then, they can provide those ranges, and the auto-scheduler can use an SMT solver to reason about the time and auxiliary memory complexities of the schedules using symbolic cost expressions that it builds for every schedule (\S~\ref{overview}).
Note that we assume that the tensors have uniform sparsities.

There are three types of constraints that we can provide the SMT solver;
\begin{enumerate}
    \item \textbf{Range constraints}: the range in which dense loop bounds and sparsities can vary (\eg line~\ref{line:rc} in Listing~\ref{lst:z3_pruning}).
    \item \textbf{Inferred constraints}: non-zero values in a sparse tensor are always less than the number of elements in its dense representation, and the non-affine loop that iterates through the sparse tensor will vary between 0 and the dense loop bound that defines the sparse tensor (\eg line~\ref{line:ic} in Listing~\ref{lst:z3_pruning}).
    \item \textbf{User-defined constraints}: other special constraints that the user may know about the loop bounds or sparsities. For instance, the user may know that one loop bound is twice of another (\eg line~\ref{line:uc} in Listing~\ref{lst:z3_pruning}).
\end{enumerate}
After providing the constraints known at compile-time, we check if one schedule is dominated by at least another schedule in terms of both iteration time and auxiliary memory complexity; if so, we remove that schedule from the Pareto frontier (See line~\ref{line:cond} in Listing~\ref{lst:z3_pruning}). 
In other words, the system removes a schedule ($s$) if there exists at least one schedule ($c$) in the schedule space such that for all possible loop bounds and sparsities, the time and memory complexities of $s$ are worse than or equal to $c$ and 
there is no set of loop bounds and sparsities for which the time and memory complexities of $s$ are better than $c$. The system does not remove a schedule if there exists at least one set values of
loop bounds and sparsities for which the time and memory complexities of $s$ are better than $c$. 
This gurantees that the system does not over-prune the schedules.

This procedure can be formally written as follows. Let user-defined constraints of loop bounds and sparsities be $\Phi$, let Z3 be the SMT solver, and the schedules from Stage 2 be $S$.
Provide $\Phi$ to Z3.
Remove $s$ from $S$ if $\exists$ $c \in S$ s.t.\\
$\exists$ loop bounds and sparsities s.t. 
\[(T(s) > T(c) \land N(s) >= N(c)) \lor (T(s) >= T(c) \land N(s) > N(c))\]
$\nexists$ loop bounds and sparsities s.t.
\[(T(s) <= T(c) \land N(s) < N(c)) \lor (T(s) < T(c) \land N(s) <= N(c))\]
Here, $T(s)$ and $N(s)$ denote the symbolic iteration time and auxiliary memory complexities of the schedule $s$, respectively.

An auto-scheduler could have this stage alone without the previous stage in Section~\ref{time_and_memory_depth_based_pruning}.
But, when any of the above conditions are $unsat$, the Z3 solver takes a long time to return.
The previous stage compares a lot more schedules than this stage.
Therefore, using a poset-based pruning strategy with absolute depth values, which is computationally efficient, is beneficial compared to using an SMT solver alone.
Nonetheless, this stage is important because we can further reduce the number of schedules evaluated at run-time with the information available at compile-time.

Furthermore, the user could use this stage alone by removing 
both the memory depth-based and poset-based pruning stages. Although 
this removes the dependency on using memory depth as a heuristic to prune the 
schedules, it takes a long time for the solver to prune the schedule and does 
not work when the number of generated schedules is large.

\subsection{Concrete Stages}\label{concrete_stages}

\heading{Time-Complexity-Based Pruning}\label{time_based_pruning}
At the first stage of filtration at run-time, we evaluate the symbolic cost expressions with real values available at the run-time and select schedules that have the least iteration time complexity such that the auxiliary memory requirement is less than 50\% of the last level cache (LLC) from the Pareto frontier.
We take 50\% as a rough margin for the selection criterion, assuming that 50\% of LLC is available for the other input and output tensors in the computation.
Multiple schedules with the same iteration time complexity can exist due to the same branched loop nest structure with different loop reorderings.
These schedules are then passed to the next stage for further pruning.

\heading{Cache-Based Pruning}\label{cache_based_pruning}
At the second stage of run-time filtration of schedules, we prune the schedules based on cache behavior. 
We have included this stage here for completeness, and it is not our primary focus. 
Many remaining schedules may share equivalent time and memory complexity due to loop reorderings, such as the loop orders $l, m$, and $m, l$ in the outer loops of Figure~\ref{fig:b-tensor-contraction-kernel}.
Since some schedules have the same time and memory complexity, if one of those schedules is not filtered away by previous stages, both of them will remain unpruned.
Thus, we have a simple model that assigns a cache access cost to each schedule. 
This cache model takes two criteria into account. 
One, it looks at the leaves in the BIG and the leaf loop index. 
If the leaf loop index is $i$ and if a tensor in the expression at that leaf branch has $i$ as the last index (\eg $B(j, i)$) or index $i$ is not present in the tensor (\eg $B(j, k)$), then the cost of access is zero. 
If $i$ is present and not in the last accessed index (\eg $B(i, j)$), then we take the cache access cost as $J$ since elements are accessed $J$ locations apart. 
We assign costs to all the leaves in the BIG and sum those to calculate a final cache access cost. 
We consider the last index of the leaves in the BIG because it has the highest impact on temporal and spatial cache locality. 
Two, we give precedence to the schedules that have the same index order as the loop order in the iteration graph. 
For instance, if tensors in the computation have $B(i,j)$ and $C(j,k)$, it would favor the loop order $i,j,k$ over $k,j,i$. 
If both these criteria are the same for two schedules, we randomly pick one of them.

\section{Evaluation}\label{evaluation}

We assess \system using a collection of sparse tensor kernels in comparison to the schedules generated by TACO~\cite{taco}. 
We compare the results of \system with Pigeon~\cite{willowahrens} and SpTTN-Cyclops~\cite{solomonik} qualitatively and quantitatively where applicable.


\paragraph{Experimental Setup} We conducted the experiments on a machine with four Non-Uniform Memory Access (NUMA) nodes of Intel(R) Xeon(R) CPU E5-4650 8-core processor (32-cores in total), operating at 2.70 GHz, with 32KB L1 data cache, 256KB L2 cache per core, and 80MB LLC shared between 4 NUMA nodes. Code compilation utilized GCC 11.4.0 with with optimization flags \texttt{-O3} \texttt{-\null-ffast-math}. The process involved a warm-up run, followed by 31 executions of the kernel computation. The results reported are the median values, accompanied by the corresponding standard deviation across the 31 runs. Parallel executions were performed on 32 threads using OpenMP.


\begin{table}[t]
\caption{Tensors and matrices used in the evaluation from various matrix and tensor collections}
\label{tab:datasets}
\begin{adjustbox}{minipage=\linewidth,scale=0.8}
\centering
\begin{center}
\begin{tabular}{lrrrr} 
    \toprule
    \textbf{Tensor} & \makecell{\textbf{Size of file on disk}}  & \textbf{Dimensions} & \textbf{Non-zeros} & \textbf{Sparsity}\\
    \midrule
    bcsstk17 & 5.4 MB & { $11K \times 11K$} & {$429K$} & 4E-3\\
    pdb1HYS & 55.0 MB & { $36K \times 36K$} & {$4.34M$} & 3E-3\\
    rma10 & 59.0 MB & { $47K \times 47K$} & {$2.37M$} & 1E-3\\
    cant & 57.0 MB & { $62K \times 62K$} & {$4.01M$} & 1E-3\\
    consph & 83.0 MB & { $83K \times 83K$} & {$6.01M$} & 9E-4\\
    cop20k\_A & 27.0 MB & { $12K \times 12K$} & {$2.62M$} & 2E-4\\
    shipsec1 & 83.0 MB & { $140K \times 140K$} & {$7.81M$} & 2E-4\\
    scircuit & 28.0 MB & { $171K \times 171K$} & {$959K$} & 3E-5\\
    mac\_econ\_fwd500 & 32.0 MB & { $207K \times 207K$} & {$1.27M$} & 9E-5\\
    webbase-1M & 68.0 MB & { $1.00M \times 1.00M$} & {$3.11M$} & 3E-6\\
    circuit5M & 2.1 GB & { $5.56M \times 5.56M$} & {$59.52M$} & 2E-6\\
    \midrule%
    vast-2015-mc1-3d & 431.0 MB & { $165K \times 11K \times 2$} & { $26.02M$} & 8.36E-08\\
    darpa1998 & 575.0 MB & { $22K \times 22K \times 23.7M$} & { $28.42M$} & 2.50E-06\\
    nell-2 & 1.5 GB & { $12K \times 9K \times 288K$} & { $76.88M$} & 5.73E-05\\
    flickr-3d & 2.6 GB & { $320K \times 2.82M \times 1.60M$} & { $112.89M$} & 3.92E-11\\
    \bottomrule
\end{tabular}
\end{center}
\end{adjustbox}
\end{table}


\paragraph{Datasets}\label{datasets} In the evaluation, we employ numerous real-world tensors sourced from the SuiteSparse Collection~\cite{suitesparse}, Network Repository~\cite{networkrepository}, Formidable Repository of Open Sparse Tensors and Tools~\cite{frosttdataset}, and the 1998 DARPA Intrusion Detection Evaluation Dataset~\cite{darpa}. The tensors and matrices used in the evaluation are shown in Table~\ref{tab:datasets}. These tensors span a wide range of sizes and sparsities. Sparse inputs to the kernels used the Compressed Sparse Fiber (CSF) format.


{\renewcommand{\arraystretch}{1.3}
\begin{table*}[t]
    \centering
    \caption{Chosen schedules after the SMT solver-based pruning stage. 
    Naming convention: ${\sparse B}$ denotes sparse tensor $B$, and ${\sparse j'}$ denotes the index of a non-affine loop.
    Different kernels inside the angle brackets denote that the fused computation can be separated into those kernels. 
    The inner branches of the loop nests are written as $\left\langle Temporary; Producer, Consumer\right\rangle$.}
    \label{fig:kernels}
    \hspace{-9em}
    \begin{adjustbox}{minipage=\linewidth,scale=0.8}
    \begin{tabular}{lll}
        \toprule
    \textbf{Kernel} & \textbf{Description} & \textbf{Chosen Schedules After Z3 Pruning}\\ [1mm]
        \midrule
    $\encircle{1}:\langle SDDMM,SpMM\rangle$ & $A_{il} = \sum_{jk} {\sparse B_{ij}}\>C_{ik}\>D_{jk}\>E_{jl}$ & $i,{\sparse j'}\left\langle t; k:t+={\sparse B_{ij}}\>C_{ik}\>D_{jk}, l:A_{il}+=t\>E_{jl} \right\rangle$ \\[1mm]
    \makecell[l]{$\encircle{2}:\langle SDDMM,SpMM,GEMM\rangle$} & $A_{im} = \sum_{jkl} {\sparse B_{ij}}\>C_{ik}\>D_{jk}\>E_{jl}\>F_{lm}$ & \makecell[l]{$i\left\langle T_{l}; \right.$\\$\enspace {\sparse j'} \left\langle t; k: t+={\sparse B_{ij}}\>C_{ik}\>D_{jk}, l:T_{l}^{1}+=t\>E_{jl} \right\rangle,$\\$\enspace\left.m,l: A_{im}+=T_{l}\>F_{lm}\right\rangle$} \\
    & & \makecell[l]{$i,{\sparse j'}\left\langle t; k:t+={\sparse B_{ij}}\>C_{ik}\>D_{jk}, m,l:A_{il}+=t\>E_{jl}\>F_{lm} \right\rangle$} \\
    & & \makecell[l]{$i,l\left\langle t; {\sparse j'},k:t+={\sparse B_{ij}}\>C_{ik}\>D_{jk}\>E_{jl}, m,l:A_{il}+=t\>F_{lm} \right\rangle$} \\ [1mm]
    $\encircle{3}:\langle SpMMH,GEMM\rangle$ & $A_{il} = \sum_{jk} {\sparse B_{ij}}\>C_{jk}\>D_{jk}\>E_{kl}$ & \makecell[l]{$i,{\sparse j'}\left\langle t; k:t+={\sparse B_{ij}}\>C_{jk}\>D_{jk}, l:A_{il}+=t\>E_{jl} \right\rangle$} \\ [1mm]
    $\encircle{4}:\langle SpMM, GEMM \rangle$ & $A_{il} = \sum_{jk} {\sparse B_{ij}}\>C_{jk}\>D_{kl}$ & \makecell[l]{$i,k\left\langle t; {\sparse j'}:t+={\sparse B_{ij}}\>C_{jk}\>, l:A_{il}+=t\>D_{kl} \right\rangle$} \\ [1mm]
    $\encircle{5}:\langle 3D\ TTMC\rangle$ & $A_{lmn} = \sum_{ijk} {\sparse B_{ijk}}\>C_{il}\>D_{jm}\>E_{kn}$ & \makecell[l]{$i,{\sparse j'},n\left\langle t; {\sparse k'}:t+={\sparse B_{ijk}}\>E_{kn}\>, m,l:A_{lmn}+=t\>C_{il}\>D_{jm} \right\rangle$} \\
    & & \makecell[l]{$i,n\left\langle T_{m}; \right.$\\$\enspace {\sparse j'}\left\langle t; {\sparse k'}:t+={\sparse B_{ijk}}\>E_{kn}, m: T_{m}+=t\>D_{jm}\right\rangle, $\\$\enspace\left.m,l:A_{lmn}+=T_{m}\>C_{il} \right\rangle$}\\
    & & \makecell[l]{$i,m,n\left\langle t; {\sparse j'},{\sparse k'}:t+={\sparse B_{ijk}}\>D_{jm}\>E_{kn}\>, m,l:A_{lmn}+=t\>C_{il} \right\rangle$} \\ [1mm]
    $\encircle{6}:\langle SpTTM,TTM\rangle$ & $A_{ilm} = \sum_{jk} {\sparse B_{ijk}}\>C_{jl}\>D_{km}$ & \makecell[l]{$i,{\sparse j'}m\left\langle t; {\sparse k'}:t+={\sparse B_{ijk}}\>D_{jk}, l:A_{il}+=t\>C_{jl} \right\rangle$} \\ [1mm]
    $\encircle{7}:\langle SpTTM,SpTTM\rangle$ & ${\sparse A_{ijm}} = \sum_{jk} {\sparse B_{ijk}}\>C_{kl}\>D_{lm}$ & \makecell[l]{$i,{\sparse j'}l\left\langle t; {\sparse k'}:t+={\sparse B_{ijk}}\>C_{kl}, m:{\sparse A_{il}}+=t\>D_{lm} \right\rangle$} \\ [1mm]
    $\encircle{8}:\langle MTTKRP,GEMM\rangle$ & $A_{im} = \sum_{jk} {\sparse B_{ikl}}\>C_{lj}\>D_{kj}\>E_{jm}$ & \makecell[l]{$i,{\sparse j'}\left\langle t; {\sparse k'}l:t+={\sparse B_{ijk}}\>C_{kl}\>D_{kj}, m:A_{il}+=t\>E_{jm} \right\rangle$} \\ [1mm]
        \bottomrule
    \end{tabular}
\end{adjustbox}
\end{table*}
}

{\renewcommand{\arraystretch}{1.0}
\begin{table}[t!]
\caption{The number of schedules after each stage in the pruning pipeline. The numbers in the parenthesis denote the number of different (time complexity, auxiliary memory complexity) pairs. Stage 3 (Skipping Stages 1 \& 2) is time-limited to 24 hours per kernel.}
\label{tab:schedule-counts}
\begin{adjustbox}{minipage=\linewidth,scale=0.85}
    \begin{center}
    \begin{tabular}{lrrrrrr}
        \toprule
        \textbf{Kernel} & \makecell{\textbf{Generated}\\\textbf{Schedules}} & \makecell{\textbf{Stage 1}\\\textbf{Mem Depth}} & \makecell{\textbf{Stage 2}\\\textbf{Depth Poset}} & \makecell{\textbf{Stage 3}\\\textbf{SMT Solver}} & \makecell{\textbf{Stage 3}\\\textbf{(Skipping}\\\textbf{Stage 2)}} & \makecell{\textbf{Stage 3}\\\textbf{(Skipping}\\\textbf{Stages 1 \& 2)}} \\
        \midrule
    $\encircle{1}:$ & \num{16169} & \num{1472} (255) & 1 (1) & 1 (1) & 1 (1) & 1 (1) \\
    $\encircle{2}:$ & \num{145448232} & \num{207129} (\num{18277}) & 224 (43) & 8 (3) & 32 (4) & timeout \\
    $\encircle{3}:$ & \num{7426} & 692 (133) & 2 (1) & 2 (1)  & 2 (1) & 2 (1) \\
    $\encircle{4}:$ & \num{258} & 128 (34) & 2 (1) & 2 (1) & 2 (1) & 2 (1) \\
    $\encircle{5}:$ & \num{14701776} & \num{30203} (\num{2541}) & 101 (26) & 16 (3) & 16 (3) & timeout \\
    $\encircle{6}:$ & \num{2561} & 352 (60) & 3 (1) & 3 (1)  & 3 (1) & 3 (1) \\
    $\encircle{7}:$ & \num{109} & 46 (13) & 1 (1) & 1 (1) & 1 (1) & 1 (1) \\
    $\encircle{8}:$ & \num{58127} & \num{4715} (728) & 5 (2) & 2 (1) & 2 (1) & timeout \\
        \bottomrule
    \end{tabular}
    \end{center}
\end{adjustbox}
\end{table}
}


{\renewcommand{\arraystretch}{0.8}
\begin{table}[t!]
\caption{Time taken for each compile-time stage. Stage 3 (Skipping Stages 1 \& 2) is time-limited to 24 hours per kernel.}
\label{tab:schedule-compile-times}
\hspace{-1.5em}
\begin{adjustbox}{minipage=\linewidth,scale=0.85}
\begin{center}
\begin{tabular}{lrrrrrr}
\toprule
\textbf{Kernel} & \makecell{\textbf{Schedule}\\\textbf{Generation}\\\textbf{(16 threads)}} & \makecell{\textbf{Stage 1}\\\textbf{Mem Depth}} & \makecell{\textbf{Stage 2}\\\textbf{Depth Poset}} & \makecell{\textbf{Stage 3}\\\textbf{SMT Solver}} & \makecell{\textbf{Stage 3}\\\textbf{(Skipping}\\\textbf{Stage 2)}} & \makecell{\textbf{Stage 3}\\\textbf{(Skipping}\\\textbf{Stages 1 \& 2)}} \\
\midrule
$\encircle{1}:$ & 631.4 ms & 13.0 ms & 1723.2 ms & 121.5 ms & 2877.4 ms & 5.6 s \\
$\encircle{2}:$ & 10740.5 s {} {} {} & 167.5 s {} {} {} & 454.1 s {} {} {} & 1.5 s {} {} {} & 1538.9 s {} {} {} & timeout \\
$\encircle{3}:$ & 475.9 ms & 3.7 ms & 896.7 ms & 48.0 ms & 2623.4 ms & 1.7 s \\
$\encircle{4}:$ & 54.6 ms & 0.5 ms & 56.6 ms & 49.7 ms & 327.0 ms & 0.8 s \\
$\encircle{5}:$ & 864.3 s {} {} {} & 71.9 s {} {} {} & 25.8 s {} {} {} & 1.5 s {} {} {} & 4567.1 s {} {} {} & timeout \\
$\encircle{6}:$ & 304.6 ms & 6.2 ms & 238.2 ms & 50.0 ms & 1244.2 ms & 5.4 s \\
$\encircle{7}:$ & 35.3 ms & 0.3 ms & 46.6 ms & 48.3 ms & 486.7 ms & 0.9 s \\
$\encircle{8}:$ & 4.2 s {} {} {} & 57.8 ms & 4.7 s {} {} {} & 123.2 ms & 204.1 s {} {} {} & timeout \\
    \bottomrule
\end{tabular}
\end{center}
\end{adjustbox}
\end{table}
}

{\renewcommand{\arraystretch}{1.0}
\begin{table}[t]
\caption{Time taken for each run-time stage including the code generation and compilation times.}
\label{tab:schedule-run-times}
\begin{adjustbox}{minipage=\linewidth,scale=0.8}
\centering
\begin{center}
\begin{tabular}{lrrrrrr}
\toprule
\multirow{3}{*}{\textbf{Kernel}} & \multicolumn{3}{c}{\textbf{Run-time Filtration}} & \multicolumn{3}{c}{\textbf{Build Time}} \\
\cmidrule(lr){2-4} \cmidrule(lr){5-7}
& \makecell{\textbf{Stage 4}\\\textbf{(us)}} & \makecell{\textbf{Stage 5}\\\textbf{(us)}} & \makecell{\textbf{Total}\\\textbf{(us)}} & \makecell{\textbf{Codegen}\\\textbf{(ms)}} & \makecell{\textbf{Compile}\\\textbf{(ms)}} & \makecell{\textbf{Total}\\\textbf{(ms)}} \\
\midrule
$\encircle{1}:$ & 185.2 &	111.1 & 296.2 &	10.9 &	163.4 &	174.3 \\
$\encircle{2}:$ & 265.1 &	94.6 &	359.7 &	12.0 &	364.2 &	376.1 \\
$\encircle{3}:$ & 159.6 &	111.3 &	270.9 &	12.1 &	174.4 &	186.5 \\
$\encircle{4}:$ & 161.6 &	125.9 &	287.5 &	12.1 &	174.6 &	186.8 \\
$\encircle{5}:$ & 220.3 &	209.4 &	429.7 &	16.5 &	251.6 &	268.1 \\
$\encircle{6}:$ & 158.0 &	115.9 &	274.0 &	14.9 &	205.2 &	220.1 \\
$\encircle{7}:$ & 198.0 &	82.3 &	280.3 &	11.6 &	189.3 &	200.8 \\
$\encircle{8}:$ & 198.0 &	101.6 &	299.6 &	15.2 &	173.0 &	188.2 \\
\bottomrule
\end{tabular}
\end{center}
\end{adjustbox}
\end{table}
}

\begin{figure*}[ht]
    \begin{adjustbox}{minipage=\linewidth,scale=0.95}
    \centering
    \begin{subfigure}[t]{0.35\textwidth}
        \centering
        \includegraphics[width=1.0\columnwidth]{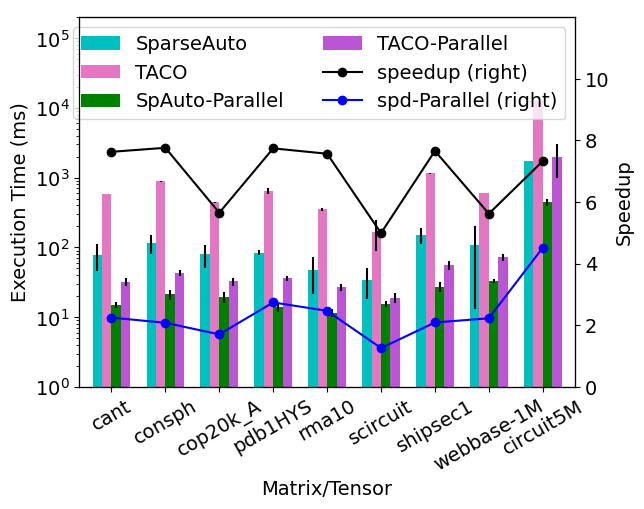}
        \vspace{-18pt}
        \caption{$\brackets{SDDMM,SpMM}$}
        \vspace{-10pt}
        \label{fig:eval1}
    \end{subfigure}%
    \hspace{20pt}
    \begin{subfigure}[t]{0.35\textwidth}
        \centering
        \includegraphics[width=1.0\columnwidth]{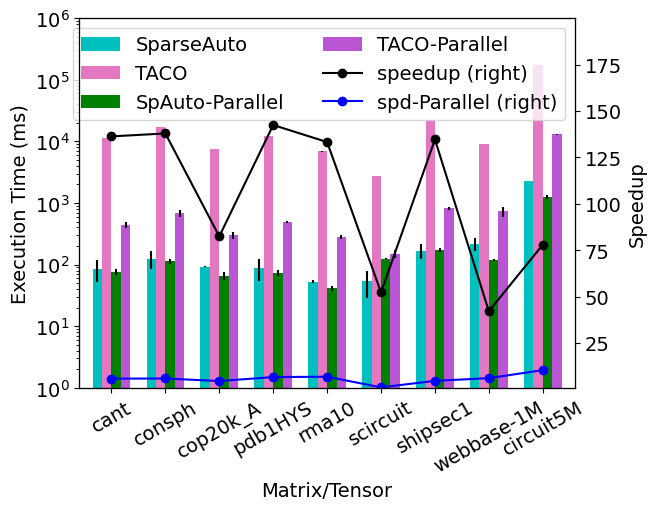}
        \vspace{-18pt}
        \caption{$\brackets{SDDMM,SpMM,GEMM}$}
        \vspace{-10pt}
        \label{fig:eval2}
    \end{subfigure}%
    \vskip\baselineskip
    \begin{subfigure}[t]{0.35\textwidth}
        \centering
        \includegraphics[width=1.0\columnwidth]{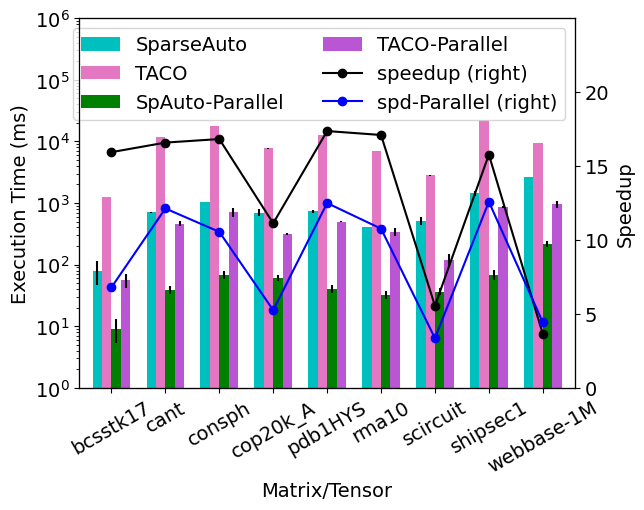}
        \vspace{-18pt}
        \caption{$\brackets{SpMMH,GEMM}$}
        \vspace{-10pt}
        \label{fig:eval3}
    \end{subfigure}%
    \hspace{20pt}
    \begin{subfigure}[t]{0.35\textwidth}
        \centering
        \includegraphics[width=1.0\columnwidth]{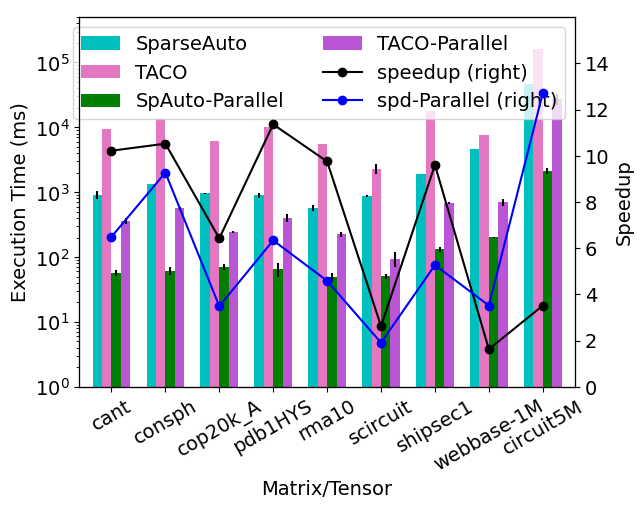}
        \vspace{-18pt}
        \caption{$\brackets{SpMM, GEMM}$}
        \vspace{-10pt}
        \label{fig:eval4}
    \end{subfigure}%
    \vskip\baselineskip
    \begin{subfigure}[t]{0.35\textwidth}
        \centering
        \includegraphics[width=1.0\columnwidth]{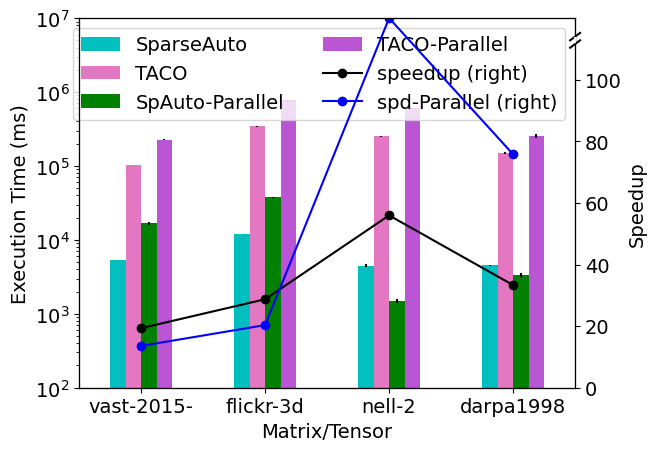}
        \vspace{-18pt}
        \caption{$\brackets{3D\ TTMC}$}
        \vspace{-10pt}
        \label{fig:eval5}
    \end{subfigure}%
    \hspace{20pt}
    \begin{subfigure}[t]{0.35\textwidth}
        \centering
        \includegraphics[width=1.0\columnwidth]{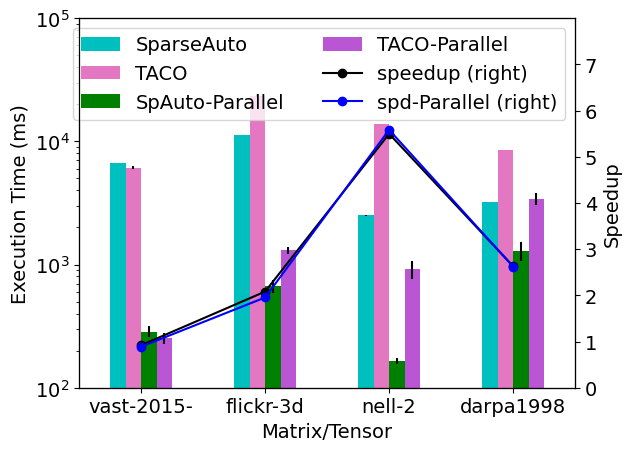}
        \vspace{-18pt}
        \caption{$\brackets{SpTTM,TTM}$}
        \vspace{-10pt}
        \label{fig:eval6}
    \end{subfigure}%
    \vskip\baselineskip
    \begin{subfigure}[t]{0.35\textwidth}
        \centering
        \includegraphics[width=1.0\columnwidth]{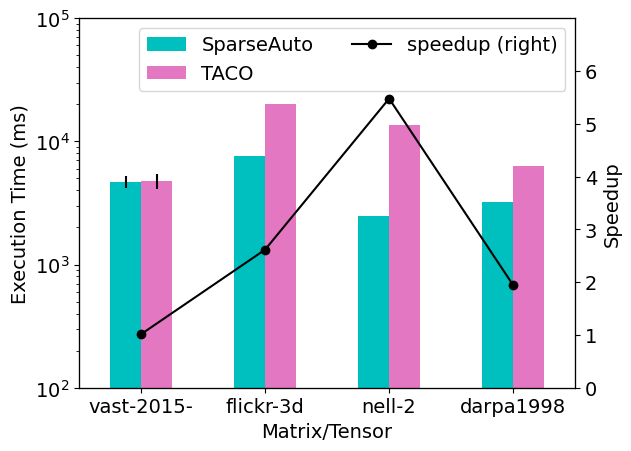}
        \vspace{-18pt}
        \caption{$\brackets{SpTTM,SpTTM}$}
        \vspace{-5pt}
        \label{fig:eva7}
    \end{subfigure}%
    \hspace{20pt}
    \begin{subfigure}[t]{0.35\textwidth}
        \centering
        \includegraphics[width=1.0\columnwidth]{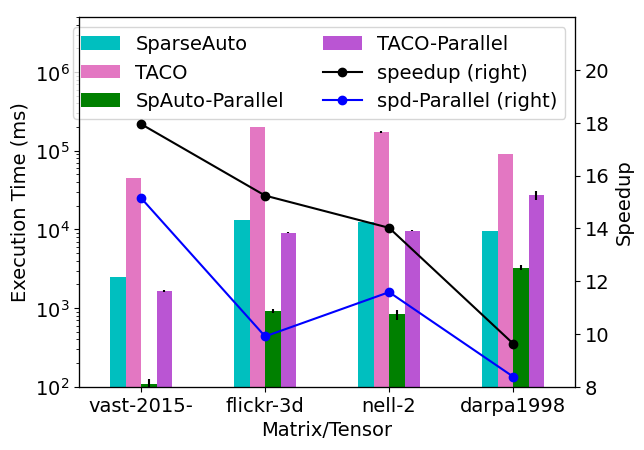}
        \vspace{-18pt}
        \caption{$\brackets{MTTKRP,GEMM}$}
        \vspace{-5pt}
        \label{fig:eval8}
    \end{subfigure}%
    \end{adjustbox}
    \caption{Performance Comparison with TACO.}
    \label{fig:evaluation}
\end{figure*}

\paragraph{Kernels}\label{kernels}
We compare the performance of \system and TACO~\cite{taco} using kernels in Table~\ref{fig:kernels}.
The kernel naming conventions are as follows: $\langle SDDMM,\allowbreak SpMM\rangle$ indicates that the kernel is a combination of $SDDMM$ and $SpMM$, and the kernel can be decomposed into these two sub-kernels, each capable of being executed sequentially.
The evaluation incorporates various combinations of the following kernels.
$SDDMM$ Sampled Dense-Dense Matrix Multiplication and $SpMM$ Sparse Matrix-Matrix Multiplication are used in graph neural networks. In this context, the $SDDMM$ operation is used in computing attention values along the edges of a graph, then $SpMM$ is used after the $SDDMM$ operation to transform the feature vector of each node, and the $GEMM$ operation is used for multiplication with a weight matrix~\cite{sparselnr}. $\langle 3D\>TTMC\rangle$ Tensor-Times Matrix Contractions are used in Tucker Decompositions~\cite{tucker}. Matrizied Tensor Times Khatri-Rao product ($MTTKRP$) is used in sparse computations such as signal processing and computer vision~\cite{mttkrp-cite}. Sparse Tensor Times Matrix ($SpTTM$) operation is used in data mining and data analytics applications and is a sub-computation in Tucker Decomposition~\cite{tucker}.

\paragraph{Number of schedules and overheads of each stage}
Table~\ref{tab:schedule-counts} shows the schedule counts of each stage in the pruning pipeline, and
Table~\ref{tab:schedule-compile-times} shows the corresponding execution times for each of these stages.
These tables expose a correlation between execution times and the number of schedules to process.
Also, Table~\ref{tab:schedule-compile-times} shows that Depth Poset-based pruning (Stage 2) helps to save on expensive SMT work (Stage 3) because the column Stage 3 (Skipping Stage 2) is always longer than Stage 2 and 3 combined.

Table~\ref{tab:schedule-run-times} shows the time taken for the two run-time stages, including the code generation and compilation times.
While we execute the codegen and compilation at run-time (for SparseAuto and TACO alike),
both can be done offline, at compile-time, as an optimization.
The optimization can work by maintaining a mapping from schedules chosen at compile-time to the corresponding compiled functions.
Using this mapping at run-time, we can lookup the code for the schedule we deem the best.
Ultimately, the time taken for the codegen and compilation at run-time is not a concern in practice.

\subsection{Performance Comparison with TACO}

Table~\ref{fig:kernels} shows the selected kernels after the compile-time pruning stages. Table~\ref{tab:schedule-counts} shows the number of schedules after each stage in the pruning pipeline. Furthermore, we bypass the second stage in the pruning pipeline and directly apply the SMT solver-based pruning in Stage 3 to the output from Stage 1. We observe that for kernel $\encircle{2}$, the number of schedules spared when Stage 2 is bypassed is 32 compared to the 8 schedules spared with Stage 2. For other kernels, the number of final schedules is the same with or without Stage 2. This indicates the effectiveness of the Depth Poset-based pruning in Stage 2. We also see that some of the schedules in Stage 2 are pruned in Stage 3, indicating the effectiveness of the SMT solver-based pruning in Stage 3.

Figure~\ref{fig:evaluation} shows the execution times and speedups of the selected schedules against the default TACO schedule. We observe orders of magnitude better performance compared to TACO. Although we do not reason about the effects of parallel execution in our auto-scheduler, we report the parallel performance of schedules by parallelizing the outer loops using OpenMP for completeness. We observe that parallel executions of the schedules have similar gains over TACO. We report only the serial execution times for $\langle SpTTM, SpTTM\rangle$ because the output of the kernel is sparse.


\subsection{Performance Comparison with Auto-Schedulers from Prior Work}

\paragraph{Comparison with Pigeon~\cite{willowahrens}.}

Pigeon introduces an auto-scheduler based on the time complexity of the schedule. They explore the search space with data layout transforms and transposes of sparse tensors, targetting an offline schedule selection. Given ${\sparse A_{ij}}$, they would consider schedules having both ${\sparse A_{ij}},$ and ${\sparse A_{ji}}$, with corresponding index orders of $i,\ j$ and $j,\ i$ in the generated schedules, whereas \system does not explore the schedules having the index order of $j,\ i$ because this requires original ${\sparse A_{ij}}$ to be transposed. We remove the schedules with transpositions in Pigeon to make a fair comparison in this study.

Considering the schedule space without transpositions, we search a larger space because of the recursive application of loop/kernel fusion/fission. Pigeon only explores one level of imperfectly nested loops. $\brackets{TTMC}$ and $\brackets{SDDMM, SpMM, GEMM}$ kernels in Table~\ref{fig:kernels} contains schedules with multiple levels of imperfect nesting. However, we could not compare the performance of these two kernels against Pigeon as their schedule generation timed out after 48 hours.

\begin{figure}[t]
    \begin{adjustbox}{minipage=\linewidth,scale=1.0}
    \centering
    \begin{subfigure}[t]{0.54\textwidth}
        \centering
        \includegraphics[width=1.0\columnwidth]{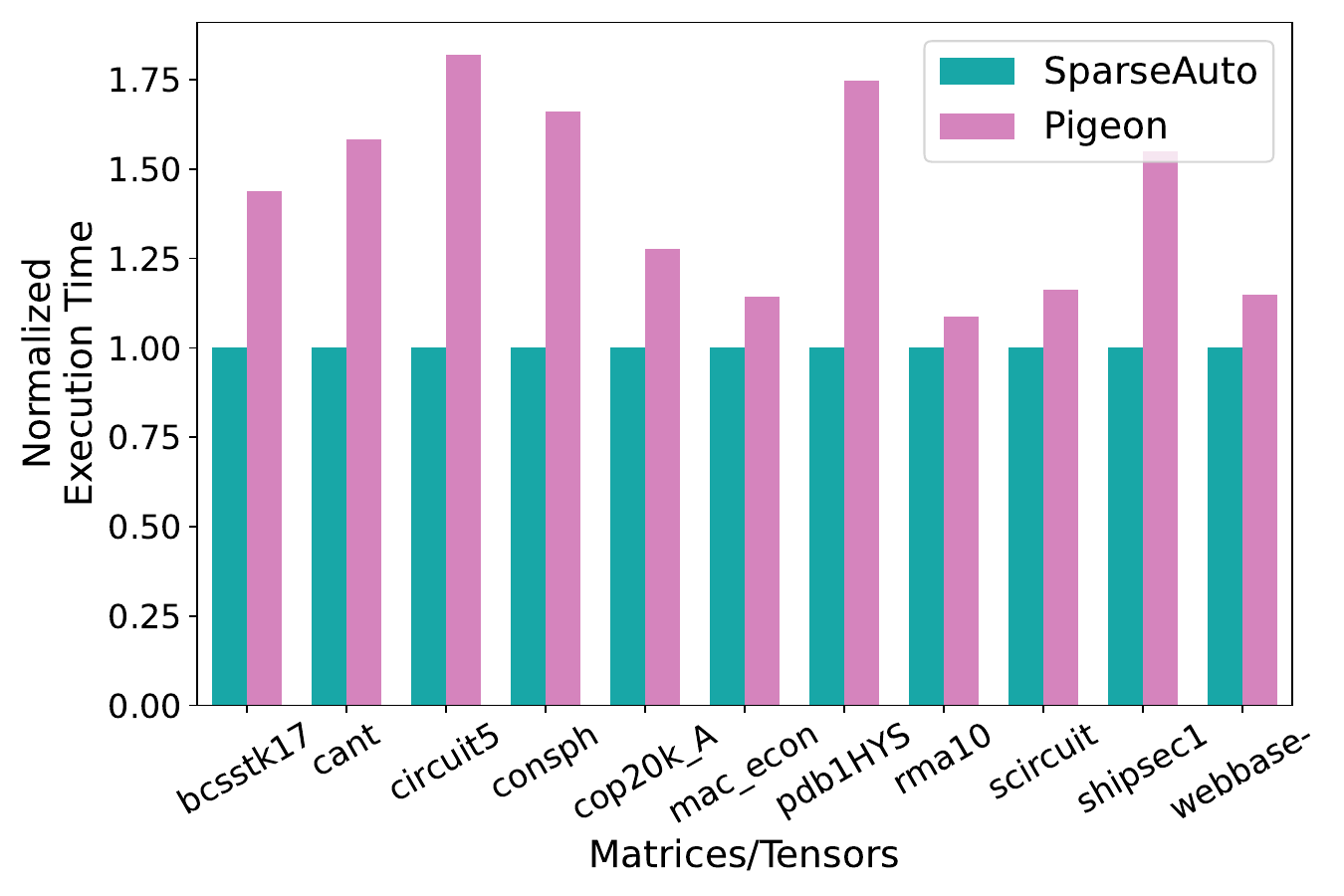}
        \vspace{-1.0em}
        \caption{}
        \vspace{-0.5em}
    \end{subfigure}%
    \hspace{2em}%
    \begin{subfigure}[t]{0.24\textwidth}
        \centering
        \includegraphics[width=1.0\columnwidth]{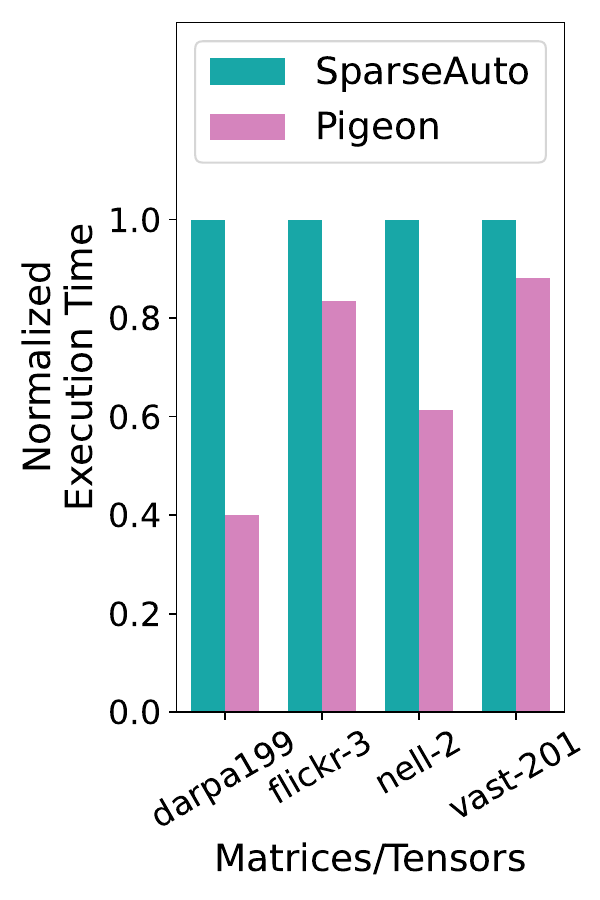}
        \vspace{-1.0em}
        \caption{}
        \vspace{-0.5em}
    \end{subfigure}%
    \end{adjustbox}
    \caption{Performance comparison of \system and Ahrens~\etal~\cite{willowahrens} for $\brackets{SpMM, GEMM}$ and $\brackets{SpTTM, SpTTM}$ kernels.}
    \label{fig:compare-ahrens}
\end{figure}

$\brackets{SDDMM, SpMM}$ kernel gave similar performance for both \system and Pigeon. We could not compare against $\brackets{SpMMH,GEMM}$, $\brackets{MTTKRP, GEMM}$, $\brackets{SpTTM, TTM}$ due to reasons such as all selected schedules that can be evaluated on TACO having data layout transforms and transposes, incorrect schedules and errors according to our experiments. We compare the performance of $\brackets{SpMM, GEMM}$, and $\brackets{SpTTM, SpTTM}$ kernels in Figure~\ref{fig:compare-ahrens}. We observe that \system outperforms Pigeon in $\brackets{SpMM, GEMM}$ kernel, and Pigeon outperforms \system in $\brackets{SpTTM, SpTTM}$ kernel. In $\brackets{SpTTM, SpTTM}$, the innermost loop is the non-affine loop in \system selected schedule. Hence, \system schedule has worse cache access patterns compared to the Pigeon schedule.

\paragraph{Comparison with SpTTN-Cyclops~\cite{solomonik}.}

\begin{figure}[t]
    \centering
    \begin{subfigure}[t]{0.38\textwidth}
        \centering
        \includegraphics[width=1.0\columnwidth]{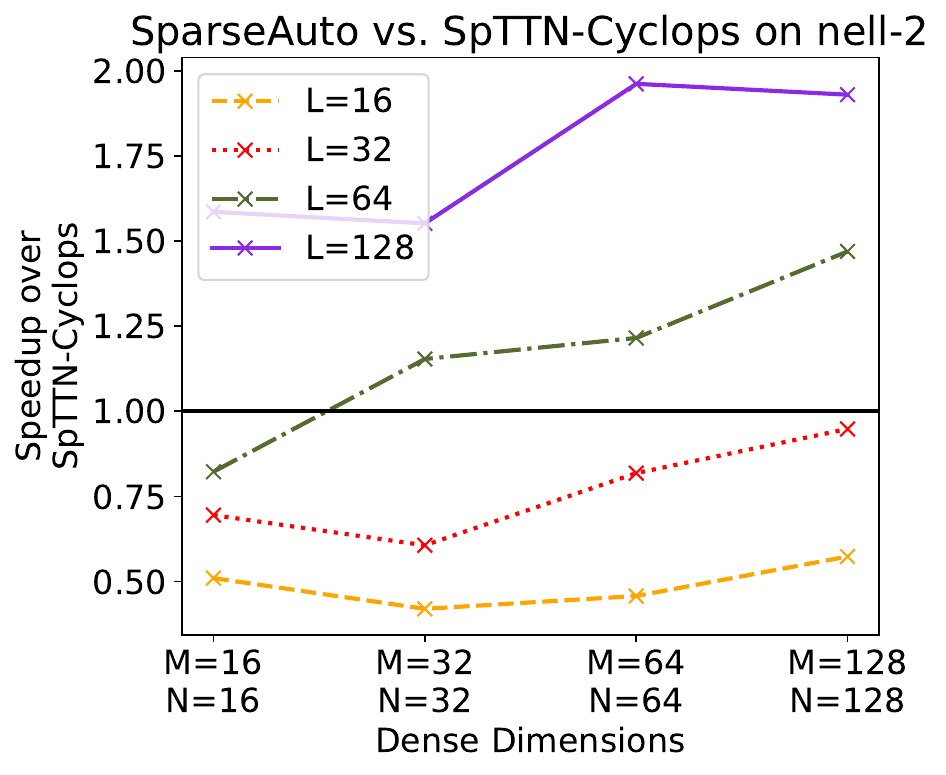}
        \vspace{-18pt}
        \caption{$\ $}
        \vspace{-8pt}
        \label{fig:cyclops-nell2}
    \end{subfigure}%
    \hspace{2em}%
    \begin{subfigure}[t]{0.38\textwidth}
        \centering
        \includegraphics[width=1.0\columnwidth]{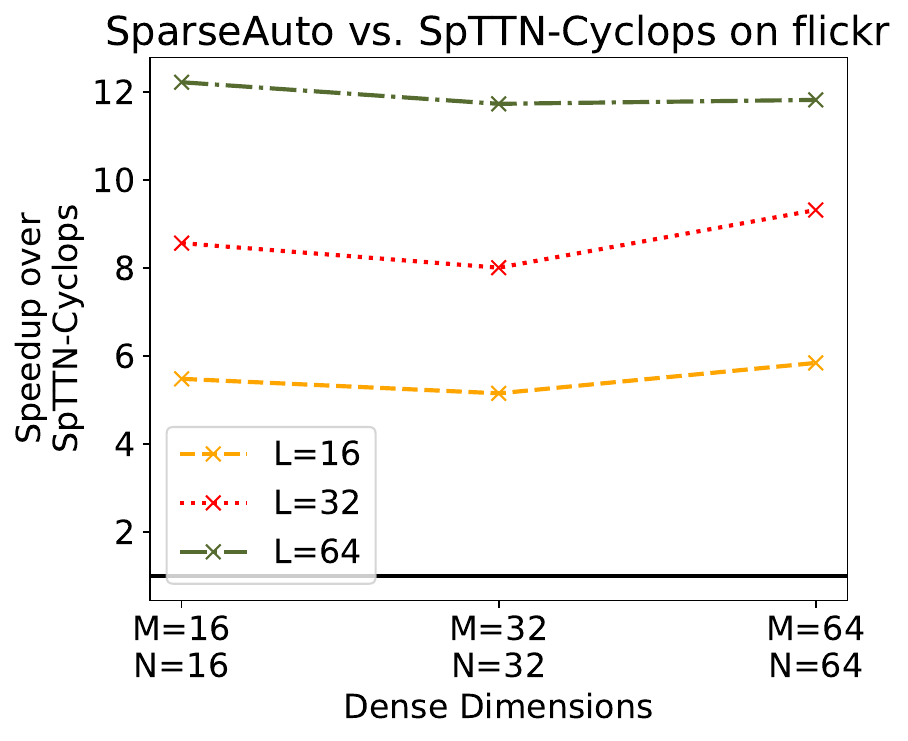}
        \vspace{-18pt}
        \caption{$\ $}
        \vspace{-8pt}
        \label{fig:cyclops-flickr}
    \end{subfigure}%
    \caption{Performance Comparison with SpTTN-Cyclops~\cite{solomonik} of $TTMC$ kernel by varying dense dimensions $L$, $M$, and $N$.}
    \label{fig:compare-cyclops}
\end{figure}

This case study shows the effect of auxiliary memory on performance. We contrast our chosen schedule with SpTTN-Cyclops's selected schedule of the $\langle 3D\>TTMC\rangle$ kernel, and evaluate on the $flickr$ and $nell-2$ datasets (refer to Figure~\ref{fig:compare-cyclops}). Their auto-scheduler, not optimized for memory, results in a schedule $i\langle T2_{mn};$$\>{\sparse j'}\langle T1_{n}; {\sparse k'},n:T1_{n}+={\sparse B_{ijk}}\>E_{kn}, m,n: T2_{mn}+=T1_{n}\>D_{jm}\rangle;$$\>l,m,n:A_{lmn}+=T2_{mn}\>C_{il}\rangle$ with one 2D and one 1D intermediate temporaries. In contrast, our schedule $i,n\langle T_{m};$$\>{\sparse j'}\langle t; {\sparse k'}:t+={\sparse B_{ijk}}\>E_{kn}, m: T_{m}+=t\>D_{jm}\rangle;$$\>m,l:A_{lmn}+=T_{m}\>C_{il}\rangle$ utilizes only one scalar and one 1D intermediate temporaries. Our schedule tends to outperform when temporary sizes (dictated by M and N dimensions) are large and temporaries are accessed more frequently (dictated by L). For smaller temporaries and fewer temporary access frequencies, SpTTN-Cyclops tends to perform better. It's important to note that both schedules share the same iteration time complexity, with speedup differences arising from cache accesses. SpTTN-Cyclops maps computations to BLAS calls, a detail omitted in our evaluation.

\subsection{Global Schedule Comparison, Scalability, and the Effect of Transposition}

\begin{figure}[t]
    \centering
    \begin{subfigure}[t]{0.38\textwidth}
        \centering
        \includegraphics[width=1.0\columnwidth]{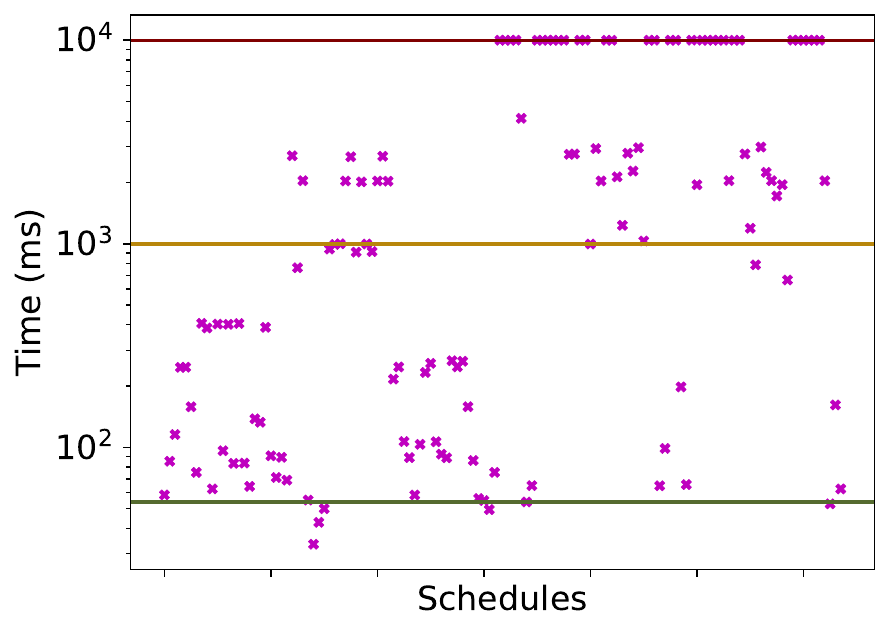}
        \vspace{-18pt}
        \caption{Dense dim $K$, $L$ set to 64}
        \vspace{-5pt}
        \label{fig:all-64}
    \end{subfigure}%
    \hspace{2em}%
    \begin{subfigure}[t]{0.38\textwidth}
        \centering
        \includegraphics[width=1.0\columnwidth]{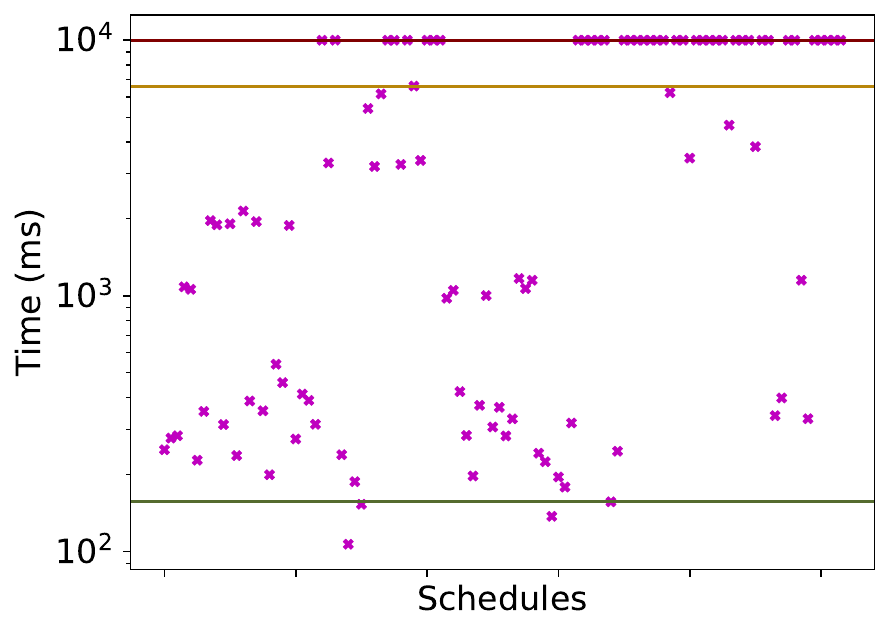}
        \vspace{-18pt}
        \caption{Dense dim $K$, $L$ set to 128}
        \vspace{-5pt}
        \label{fig:all-128}
    \end{subfigure}%
    \caption{Performance Comparison of \system with all of the other schedules for $\langle SpMM, GEMM\rangle$ using bcsstk17. The bottom, middle, and top lines correspond to the execution times of the \system, the default TACO schedule, and the timeout.}
    \label{fig:global-eval}
\end{figure}

\paragraph{Comparison of performance against all schedules}
We assess the performance of our chosen schedule in comparison to all other schedules illustrated in Figure~\ref{fig:global-eval} for a kernel with fewer schedule options, and for a smaller matrix. \system-selected schedule emerges as one of the top-performing schedules.

\begin{figure}[t]
    \vspace{-0.5em}
    \centering
    \begin{subfigure}[t]{0.38\textwidth}
        \centering
        \includegraphics[width=1.0\columnwidth]{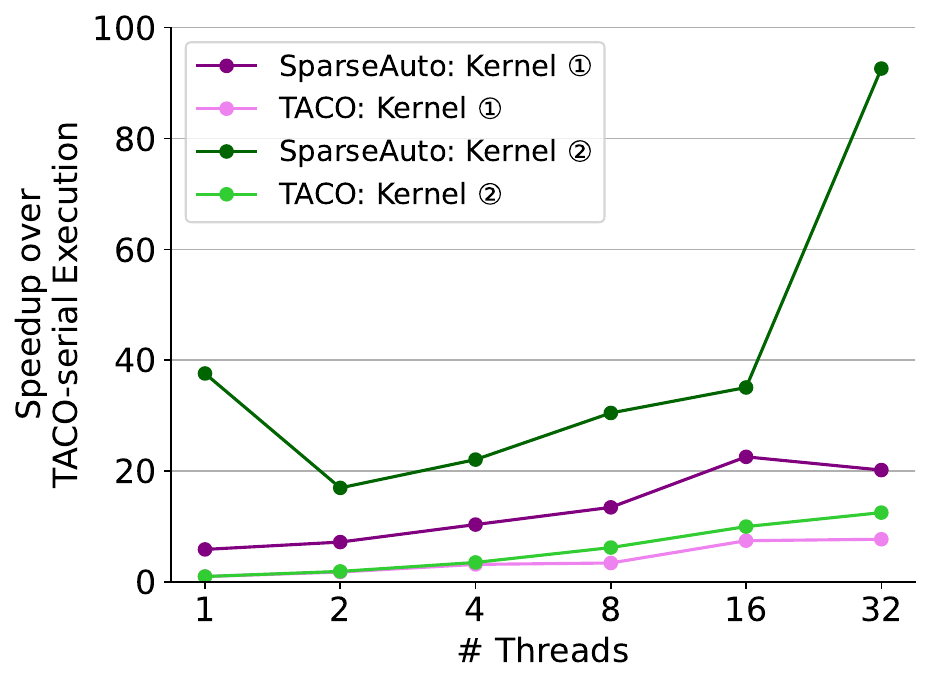}
        \vspace{-15pt}
        \caption{$\ $}
        \vspace{-5pt}
        \label{fig:webbase-scaling}
    \end{subfigure}%
    \hspace{2em}%
    \begin{subfigure}[t]{0.38\textwidth}
        \centering
        \includegraphics[width=1.0\columnwidth]{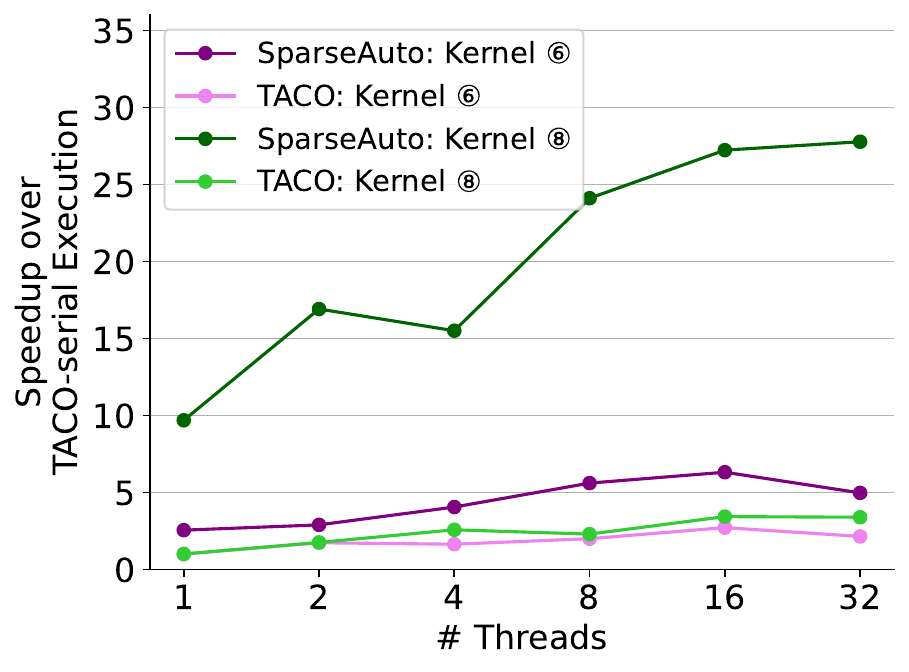}
        \vspace{-15pt}
        \caption{$\ $}
        \vspace{-5pt}
        \label{fig:darpa-scaling}
    \end{subfigure}%
    \caption{Scaling of the \system w.r.t. TACO serial.}
    \label{fig:scaling-eval}
    \vspace{-1.2em}
\end{figure}

\paragraph{Scalability}
We report the scalability results for completeness even though our auto-scheduler does not consider parallelization. 
The \system-selected schedules scale comparably to schedules generated by TACO, see Figure~\ref{fig:scaling-eval}.
Figure~\ref{fig:webbase-scaling} and Figure~\ref{fig:darpa-scaling} are generated using Webbase-1M and darpa1998, respectively. 
Scaling is weak in cases with small dense dimensions (\eg 16) but improves with larger dense dimensions.

\paragraph{Effect of transpositions on performance}
SparseLNR~\cite{sparselnr} selects the schedule $i\langle T_{k}{,}$$\ {\sparse j'}k:T_{k}+={\sparse B_{ij}}\>C_{jk}\>{,}$$\ lk:A_{il}+=T_{k}\>D_{lk} \rangle$ with an additional 1D auxiliary memory for the kernel $\brackets{SpMM, GEMM}$. Notice that the transposed $D_{kl}$ was used here as opposed to $D_{lk}$ in our schedule.
\system chosen schedule surpasses SparseLNR in terms of auxiliary memory efficiency, as our schedule in Table~\ref{fig:kernels} utilizes only a single extra scalar memory. We transpose $C_{jk}$ and $D_{kl}$ and evaluate against SparseLNR.
While SparseLNR schedule with 1D auxiliary memory outperforms ours in cases like $A_{il} = \sum_{jk} {\sparse B_{ij}}\>C_{jk}\>D_{lk}$, and $A_{il} = \sum_{jk} {\sparse B_{ij}}\>C_{jk}\>D_{kl}$, \system schedule with scalar auxiliary memory excels in instances like $A_{il} = \sum_{jk} {\sparse B_{ij}}\>C_{kj}\>D_{kl}$, and $A_{il} = \sum_{jk} {\sparse B_{ij}}\>C_{kj}\>D_{lk}$. 
The discrepancy arises from cache access effects with transposed matrices.
Addressing cache misses and access patterns would require the auto-scheduler to delve into intricate details, surpassing the scope of this paper. We identify this as potential future work.

\subsection{Effect of Auxiliary Memory on Performance}

{\renewcommand{\arraystretch}{1.4}
\begin{table*}[t]
    \centering
    \caption{Auxiliary memory requirements for different kernels. The right four columns show the number of elements required to store the intermediate results. The \lq---\rq\ label shows that the given schedule is not available with the given framework.
    }
    \label{fig:memory-requirements}
    \hspace{-5.2em}
    \begin{adjustbox}{minipage=\linewidth,scale=0.8}
    \begin{tabular}{llrrrr}
        \toprule
    \multirow{2}{*}{\textbf{Kernel}} & \multirow{2}{*}{\textbf{Description}} & \multicolumn{4}{c}{\textbf{Auxiliary Memory Requirements}} \\
     \cmidrule{3-6}
    & & \textbf{TACO} & \makecell{\textbf{Sparse-}\\\textbf{Auto}} & \makecell{\textbf{SpTTN-}\\\textbf{Cyclops}} & \textbf{Pigeon}\\ [1mm]
        \midrule
    $\encircle{1}:\langle SDDMM,SpMM\rangle$ & $A_{il} = \sum_{jk} {\sparse B_{ij}}\>C_{ik}\>D_{jk}\>E_{jl}$ & $0$ & $1$ & -- & -- \\[1mm]
    \makecell[l]{$\encircle{2}:\langle SDDMM,SpMM,GEMM\rangle$} & $A_{im} = \sum_{jkl} {\sparse B_{ij}}\>C_{ik}\>D_{jk}\>E_{jl}\>F_{lm}$ & 0 & $1$ to $1+L$ & -- & -- \\
    $\encircle{3}:\langle SpMMH,GEMM\rangle$ & $A_{il} = \sum_{jk} {\sparse B_{ij}}\>C_{jk}\>D_{jk}\>E_{kl}$ & $0$ & $1$ & -- & -- \\ [1mm]
    $\encircle{4}:\langle SpMM, GEMM \rangle$ & $A_{il} = \sum_{jk} {\sparse B_{ij}}\>C_{jk}\>D_{kl}$ & $0$ & $1$ & -- & $K$ \\ [1mm]
    $\encircle{5}:\langle 3D\ TTMC\rangle$ & $A_{lmn} = \sum_{ijk} {\sparse B_{ijk}}\>C_{il}\>D_{jm}\>E_{kn}$ & $0$ & $1$ to $1 + M$ & $N + M*N$ & -- \\ [1mm]
    $\encircle{6}:\langle SpTTM,TTM\rangle$ & $A_{ilm} = \sum_{jk} {\sparse B_{ijk}}\>C_{jl}\>D_{km}$ & $0$ & $1$ & -- & -- \\ [1mm]
    $\encircle{7}:\langle SpTTM,SpTTM\rangle$ & ${\sparse A_{ijm}} = \sum_{jk} {\sparse B_{ijk}}\>C_{kl}\>D_{lm}$ & $0$ & $1$ & -- & $L$ \\ [1mm]
    $\encircle{8}:\langle MTTKRP,GEMM\rangle$ & $A_{im} = \sum_{jk} {\sparse B_{ikl}}\>C_{lj}\>D_{kj}\>E_{jm}$ & $0$ & $1$ & -- & -- \\ [1mm]
        \bottomrule
    \end{tabular}
\end{adjustbox}
\end{table*}
}

{\renewcommand{\arraystretch}{1.0}
\begin{table}[t]
\caption{Performance with changing auxiliary memory sizes of \texttt{TTMC} kernel. 
We do not evaluate the instances marked with \lq---\rq\ due to very long execution times (\ie timeout).}
\label{tab:ttmc-auxiliary-memory}
\hspace{-3em}
\begin{adjustbox}{minipage=\linewidth,scale=0.8}
\centering
\begin{center}
\begin{tabular}{lrrrrrrrr}
\toprule
\multirow{2}{*}{\makecell{\textbf{Dense Dims}\\\textbf{L, M, N}}} & \multicolumn{2}{c}{\textbf{Aux. Mem. (B)}} & \multicolumn{3}{c}{\textbf{nell-2}} & \multicolumn{3}{c}{\textbf{flickr}} \\
\cmidrule(lr){2-3} \cmidrule(lr){4-6} \cmidrule(lr){7-9}
& \textbf{SpAuto} & \textbf{Cyclops} & \makecell{\textbf{SpAuto}\\\textbf{Time (s)}} & \makecell{\textbf{Cyclops}\\\textbf{Time (s)}} & \textbf{Speedup} & \makecell{\textbf{SpAuto}\\\textbf{Time (s)}} & \makecell{\textbf{Cyclops}\\\textbf{Time (s)}} & \textbf{Speedup} \\
\midrule
16, 16, 16 & 68 & 1088 & 4.4 & 2.2 & 0.5x & 12.3 & 67.4 & 5.5x \\
32, 16, 16 & 68 & 1088 & 4.3 & 3.0 & 0.7x & 15.2 & 130.1 & 8.6x \\
64, 16, 16 & 68 & 1088 & 5.6 & 4.6 & 0.8x & 20.9 & 254.8 & 12.2x \\
128, 16, 16 & 68 & 1088 & 5.2 & 8.2 & 1.6x & 33.0 & 542.7 & 16.4x \\
16, 32, 32 & 132 & 4224 & 13.4 & 5.6 & 0.4x & 40.6 & 208.8 & 5.2x \\
32, 32, 32 & 132 & 4224 & 13.8 & 8.3 & 0.6x & 54.3 & 434.5 & 8.0x \\
64, 32, 32 & 132 & 4224 & 15.1 & 17.5 & 1.2x & 88.9 & 1042.7 & 11.7x \\
128, 32, 32 & 132 & 4224 & 17.6 & 27.4 & 1.6x & 152.6 & 2063.8 & 13.5x \\
16, 64, 64 & 260 & 16640 & 36.3 & 16.6 & 0.5x & 170.7 & 996.9 & 5.8x \\
32, 64, 64 & 260 & 16640 & 38.9 & 31.8 & 0.8x & 213.6 & 1990.0 & 9.3x \\
64, 64, 64 & 260 & 16640 & 43.4 & 52.8 & 1.2x & 334.4 & 3952.8 & 11.8x \\
128, 64, 64 & 260 & 16640 & 52.6 & 103.3 & 2.0x & --- & --- & --- \\
16, 128, 128 & 516 & 66048 & 97.5 & 55.9 & 0.6x & --- & --- & --- \\
32, 128, 128 & 516 & 66048 & 108.6 & 102.9 & 1.0x & --- & --- & --- \\
64, 128, 128 & 516 & 66048 & 136.6 & 200.8 & 1.5x & --- & --- & --- \\
128, 128, 128 & 516 & 66048 & 221.1 & 426.9 & 1.9x & --- & --- & --- \\
\bottomrule
\end{tabular}
\end{center}
\end{adjustbox}
\end{table}
}

This section analyses the auxiliary memory requirements of the selected schedules in each of the different frameworks. 
The auxiliary memory requirements are shown in Table~\ref{fig:memory-requirements}. 
The default TACO schedule does not use any auxiliary memory because it generates perfectly nested loops (\ie linear). 
Schedules generated by SparseAuto require auxiliary memory no more than the other frameworks except the default TACO schedules.

Table~\ref{tab:ttmc-auxiliary-memory} shows the performance of SparseAuto vs. SpTTN-Cyclops when the dense dimension sizes are varied in the kernels.
Generally, the SparseAuto schedule outperforms the SpTTN-Cyclops schedule when the dense dimension bounds are higher.
Notably, the dimension bound $L$ dictates the number of times the auxiliary memory is passed between the producer and consumer.
The performance varies between the two schedules more when the value of $L$ is higher.
Furthermore, doubling $M$ and $N$ quadruples the auxiliary memory in the SpTTN-Cyclops schedule, making it less efficient than the SparseAuto schedule for larger dense dimension bounds.

\begin{figure}[t]
    \centering
    \begin{subfigure}[t]{0.38\textwidth}
        \centering
        \includegraphics[width=1.0\columnwidth]{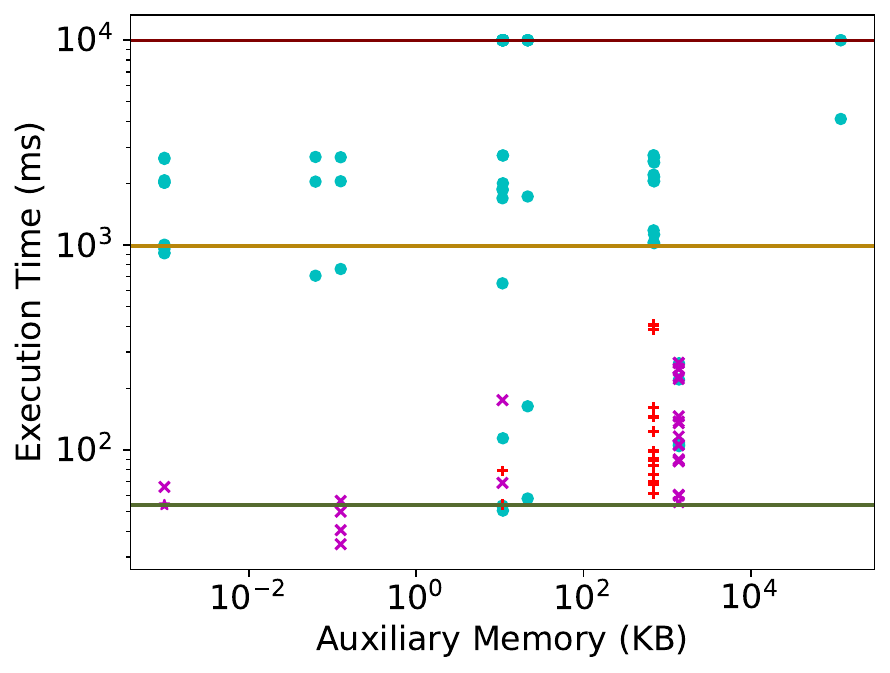}
        \vspace{-18pt}
        \caption{Dense dim $K$, $L$ set to 64}
        \vspace{-5pt}
        \label{fig:all-64-memory}
    \end{subfigure}%
    \hspace{2em}%
    \begin{subfigure}[t]{0.38\textwidth}
        \centering
        \includegraphics[width=1.0\columnwidth]{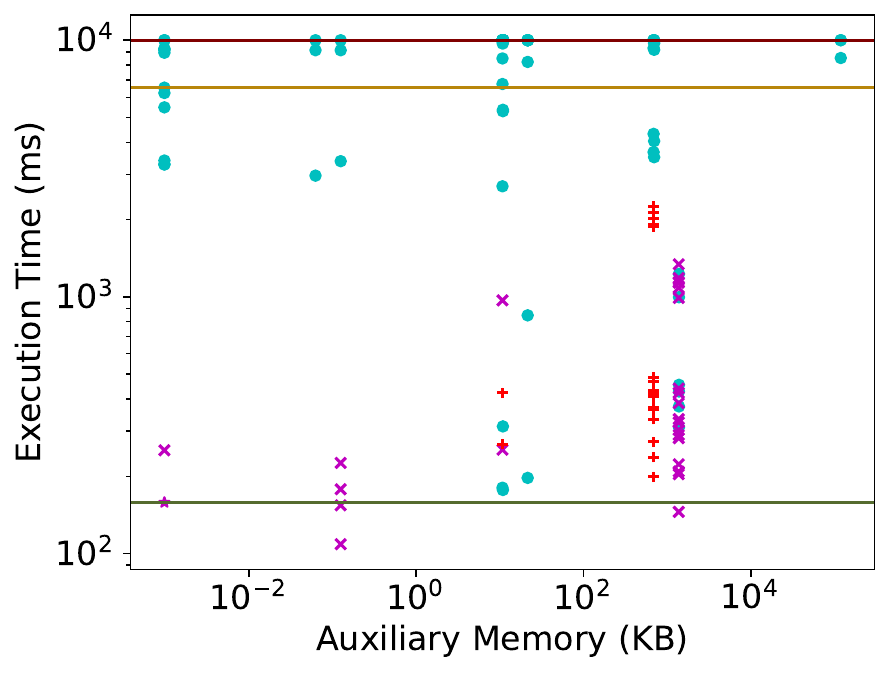}
        \vspace{-18pt}
        \caption{Dense dim $K$, $L$ set to 128}
        \vspace{-5pt}
        \label{fig:all-128-memory}
    \end{subfigure}%
    \caption{
    Auxiliary Memory Usage vs. Execution Time for $\langle SpMM,GEMM\rangle$ kernel. 
    The schedule chosen by SparseAuto is marked with a \lq$*$\rq\ in magenta. 
    The schedules with the same time complexity as the chosen schedule are shown with \lq$\times$\rq\ markers in magenta. 
    The schedules with the lowest time complexity are shown with \lq$+$\rq\ markers in red. 
    The dot markers in cyan denote the other schedules. 
    The bottom, middle and top horizontal lines mark the execution times of the schedule chosen by SparseAuto, default TACO, and the timeout limit, respectively.
    }
    \label{fig:global-eval-memory}
    \vspace{-1.0em}
\end{figure}

Figure~\ref{fig:global-eval-memory} shows that the chosen schedule uses the lowest auxiliary memory compared to the other schedules while delivering top performance. 
Many schedules have the same time complexity units as the schedule chosen by SparseAuto, but they have different sizes of auxiliary memory. 
By considering both time and memory complexities together, we can see that the number of schedules evaluated at run-time is reduced significantly. 
The schedules having the lowest time complexity (shown in red \lq$+$\rq\ markers) use much more auxiliary memory than most other schedules.

\begin{figure}[t]
    \centering
    \begin{subfigure}[t]{0.38\textwidth}
        \centering
        \includegraphics[width=1.0\columnwidth]{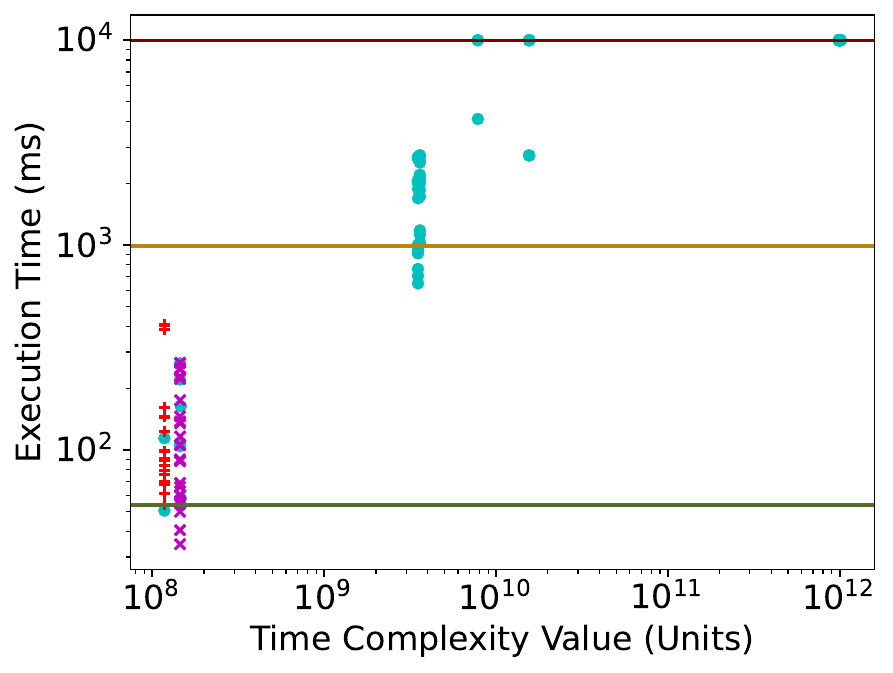}
        \vspace{-18pt}
        \caption{Dense dim $K$, $L$ set to 64}
        \vspace{-5pt}
        \label{fig:all-64-time}
    \end{subfigure}%
    \hspace{2em}%
    \begin{subfigure}[t]{0.38\textwidth}
        \centering
        \includegraphics[width=1.0\columnwidth]{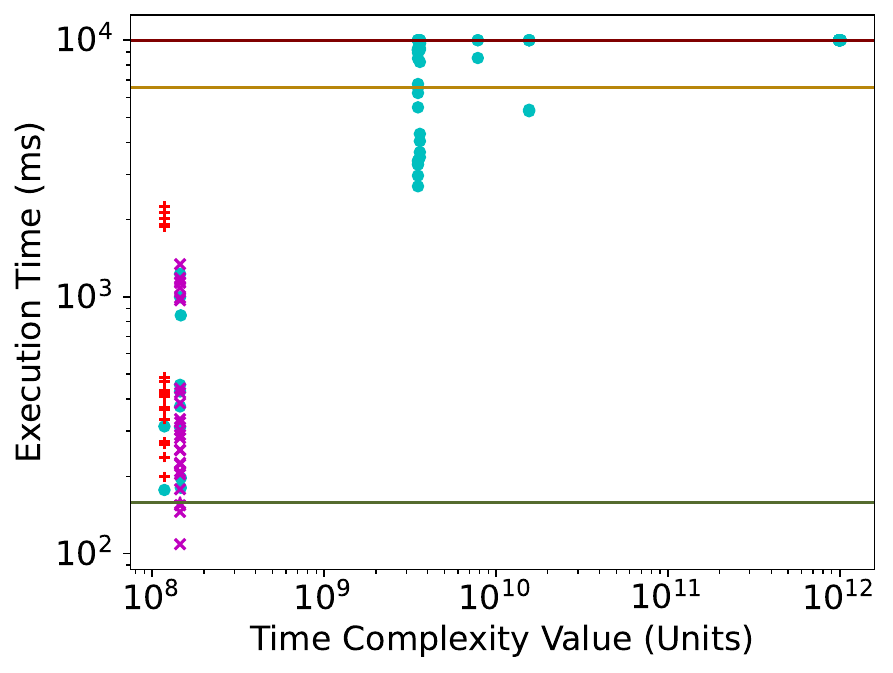}
        \vspace{-18pt}
        \caption{Dense dim $K$, $L$ set to 128}
        \vspace{-5pt}
        \label{fig:all-128-time}
    \end{subfigure}%
    \caption{
    Calculated Time Complexity vs. Execution Time for $\langle SpMM,GEMM\rangle$ kernel. 
    Schedules marked by \lq$+$\rq\ has the lowest calculated time complexity. 
    The schedules having the same memory complexity as the chosen schedule are shown as \lq$\times$\rq\ markers. 
    The schedule chosen by SparseAuto is marked with a \lq$*$\rq. 
    The bottom, middle and top horizontal lines mark the execution times of the schedule chosen by SparseAuto, default TACO, and the timeout limit, respectively.
    }
    \label{fig:global-eval-time}
    \vspace{-1.0em}
\end{figure}

Figure~\ref{fig:global-eval-time} shows the same set of schedules as Figure~\ref{fig:global-eval-memory} but using concrete time complexity on the X-axis.
Since there are many schedules with the lowest time complexity, if the system were to keep those schedules during the compile-time, a larger number of schedules would be evaluated at run-time, which would increase overhead.
The schedules with the lowest time complexity perform worse than those chosen by SparseAuto. 
This experiment shows that auxiliary memory usage is important to consider in addition to time complexity to select the best schedule.

\subsection{Discussion on the Effect of Transposition}

Our framework assumes that transposed versions of sparse tensors are not 
available at run-time. 
Hence, we only consider loop orders that do not violate the original tensors' sparse access constraints. 
For example, considering the kernel $\langle SDDMM, SpMM\rangle$, $A_{il} = \sum_{jk} {\sparse B_{ij}}\>C_{ik}\>D_{jk}\>E_{jl}$, we only consider the loop orders that have $i$ and $j$ in that order when the inner computation contains the sparse matrix ${\sparse B_{ij}}$. 
However, if the transposed version of ${\sparse B_{ij}}$, which is ${\sparse B_{ji}}$, is available at run-time, the loop orders with $j$ and $i$ in that order can also be considered.
This is a limitation of our framework.

A more advanced version of our auto-scheduler could consider the time and auxiliary memory it takes to transpose the sparse tensors and include it in the symbolic complexities of schedules. 
The transposition cost could depend on the sparse tensor format as well as the transposition algorithm. 
Taking a step further, the auto-scheduler could even consider transposing the dense tensors. 
However, this would require a more complex framework that can reason about the effects of transpositions on the performance of the schedules. 
This is a potential future work.

\section{Related Work}\label{related_works}


\heading{Sparse tensor contractions and compilers} Various works have delved into dense and sparse tensor contractions within the domains of tensor compilers, compiler optimizations, and targeted optimizations for specific kernels.

For dense tensor algebra, extensive research~\cite{saday2,saday3,saday1} has focused on CPU optimizations in memory-constrained environments. Dense transformations are comparatively straightforward due to the absence of structural or sparse access pattern constraints. GPU optimizations for dense tensor computations have been explored in works like~\cite{kim,ABDELFATTAH2016108,efficientTCE}. However, these approaches do not apply to sparse tensor computations, given the challenges posed by the intricacies of sparse data structures due to not having random access and non-affine loops.

The Tensor Contraction Engine (TCE)\cite{tce2} addresses dense tensor computations by generating code to fit available memory, utilizing a feedback loop to minimize memory. Our approach, in contrast, relies on a poset-based mechanism. Other works\cite{Lam1997OnOA,saday4} adopt similar optimizations for dense tensors. Johnnie et al.~\cite{Gray_2021} approach the tensor contraction problem as a graph problem, emphasizing dense tensor contractions.

The Sparse Polyhedral Framework~\cite{LaMielle2010EnablingCG,sparse_polyhedral_framework,sparse_polyhedral_framework2} utilizes an inspector-executor strategy to transform the data layout and schedule sparse computations, aiming to enhance both locality and parallelism. Athena~\cite{athena} and Sparta~\cite{sparta} are methodologies that offer highly optimized kernels for sparse tensor operations and contraction sequences. However, they lack support for recursive loop nest restructuring for arbitrary sparse tensor expressions.

General sparse tensor algebra compilers such as TACO~\cite{taco}, COMET~\cite{compiler_in_mlir}, and Sparsifier~\cite{aart22} in MLIR, focus on generating code for sparse tensor computations. However, they lack support for nested multiple levels of loop branches and mechanisms for exploring the search space. SparseTIR~\cite{ye2023sparsetir}, SparseLNR~\cite{sparselnr}, and ReACT~\cite{react} can generate fused sparse loops but are limited in supporting arbitrary loop nests with multiple branches and exploring the search space.

\heading{Auto-schedulers for sparse tensor contractions} Pigeon~\cite{willowahrens} introduce an auto-scheduler emphasizing loop depth and later utilize an asymptotic cost model for search space pruning. Their system does not optimize for both time and auxiliary memory complexities, operates completely offline, executes multiple schedules at the last stage, in other words they design the system for complete offline schedule selection, their search space exploration algorithm does not explore the schedules with multi-level branch nests, and lacks the use of user-defined constraints at compile time for search space pruning. Although they introduce a good cost model, their system has the disadvantages described in Section~\ref{overview}. Their framework, evaluated on TACO, faces limitations in supporting intermediate temporaries with more than one dimension and includes schedules with data layout transformations in their auto-scheduler.

SpTTN-Cyclops~\cite{solomonik} presents another auto-scheduler for sparse tensor contractions, offering a fully automated framework without user intervention in schedule selection. Unlike our approach, they do not emphasize poset-based pruning and opt for minimum loop depth schedules and then a maximum number of dense loops, which may not always be optimal, as discussed in Section~\ref{overview}. They lack support for user-defined constraints using an SMT solver to analyze schedule complexities for search space pruning. Their run-time loop generation algorithm affects evaluation time, while our method minimizes the number of schedules evaluated during run time.

\section{Discussion and Conclusion}\label{conclusion}

Auto-scheduling is a challenging problem due to the vast number of potential schedules available for a given computation—ranging from thousands to hundreds of thousands. 
Factors such as time complexity, memory usage, cache behavior, and parallelism must all be considered. 
Most systems rely on heuristic-based approaches or empirical evaluations to identify optimal schedules. 
We advocate for a systematic approach that entails dedicating considerable time to offline schedule generation and analysis. 
By investing hours in this process, most schedules can be eliminated, leaving only a select few to be evaluated at run-time. 
We propose implementing "Scheduling as a Service (SaaS)" for computations, particularly scientific workloads, which would lead to faster compute times and more efficient memory/resource utilization.

We have introduced \system, a framework for recursive loop nest restructuring, providing scheduling language support for sparse tensor contractions. 
An auto-scheduler for sparse tensor contractions was also implemented, leveraging the defined scheduling language to generate schedules. 
Among the numerous factors influencing schedule performance, we focus on two machine-independent criteria: time complexity and auxiliary memory usage, arising from variations in loop structures within sparse tensor contractions. 
\system employs a poset-based approach to prune the search space and utilizes an SMT solver for analyzing the symbolic cost of a schedule. 
Our findings demonstrate that \system delivers noteworthy performance enhancements.

\begin{acks}
We would like to thank Charitha Saumya for the valuable discussions we had regarding the \system.
This work was supported in part by the \grantsponsor{nsf}{National Science Foundation}{} awards \grantnum{nsf}{CCF-2216978}, \grantnum{nsf}{CCF-1919197} and \grantnum{nsf}{CCF-1908504}.
Any opinions, findings, and conclusions or recommendations expressed in this paper are those of the authors and do not necessarily reflect the views of the \grantsponsor{nsf}{National Science Foundation}{}.
\end{acks}

\bibliographystyle{ACM-Reference-Format}
\bibliography{bibfile}

\appendix

\end{document}